\documentclass[nopreprintline,preprint,12pt]{elsarticle}

\usepackage{amsmath}
\usepackage{amssymb}
\usepackage{lineno}
\usepackage{epsfig}
\usepackage{graphicx}
\usepackage{subcaption}
\usepackage{capt-of}
\usepackage{multirow}
\usepackage{adjustbox}
\usepackage{gensymb}
\usepackage{booktabs}
\usepackage[thinlines]{easytable}
\usepackage{mfirstuc}
\usepackage{array}
\usepackage{color}
\usepackage{soul}
\usepackage[toc,page,title]{appendix}

\newcolumntype{?}{!{\vrule width 1pt}}
\newcolumntype{M}[1]{>{\centering\arraybackslash}m{#1}}
\renewcommand{\arraystretch}{1.2}
\newcommand{\ra}[1]{\renewcommand{\arraystretch}{#1}}

\newcommand{\norm}[1]{\left\lVert#1\right\rVert}
\newcommand{\R}{\mathbb{R}}
\newcommand{\Var}{\mathrm{Var}}

\usepackage{acronym}
\acrodef{bo}[BayesOpt]{Bayesian Optimization}
\acrodef{gp}[GP]{Gaussian Process}
\acrodefplural{gp}{Gaussian Processes}
\acrodef{umap}[UMAP]{Uniform Manifold Approximation and Projection}
\acrodef{ai}[AI]{artificial intelligence}
\acrodef{rl}[RL]{reinforcement learning}
\acrodef{sgd}[SGD]{stochastic gradient descent}
\acrodef{mc}[MC]{Monte Carlo}
\acrodef{ei}[qEI]{expected improvement}
\acrodef{nei}[qNoisyEI]{noisy expected improvement}

\newcommand{\NS}{\emph{NeuroSwarms}}

\journal{Neurocomputing}

\begin{document}

\begin{frontmatter}



\title{Bayesian optimization of distributed neurodynamical controller models for
spatial navigation}


\author[apl]{Armin~Hadzic\corref{cor1}}
\author[apl,kavli]{Grace~M.~Hwang\fnref{nsfdisclaim}}
\author[jhmi]{Kechen~Zhang}
\author[apl]{Kevin~M.~Schultz}
\author[jhmi]{Joseph~D.~Monaco}

\affiliation[apl]{organization={The Johns Hopkins University/Applied Physics
Laboratory},
            city={Laurel},
            postcode={20723},
            state={MD},
            country={USA}}

\affiliation[kavli]{organization={Kavli Neuroscience Discovery Institute, Johns
Hopkins University},
            city={Baltimore},
            postcode={21218},
            state={VA},
            country={USA}}

\affiliation[jhmi]{organization={Department of Biomedical Engineering, Johns
Hopkins University School of Medicine},
            city={Baltimore},
            postcode={21205},
            state={MD},
            country={USA}}

\cortext[cor1]{arminhadzic@outlook.com}

\fntext[nsfdisclaim]{This material is based on work supported by (while serving
at) the National Science Foundation. Any opinion, findings, and conclusions or
recommendations expressed in this material are those of the authors and do not
necessarily reflect the views of the National Science Foundation.}

\begin{abstract}
Dynamical systems models for controlling multi-agent swarms have demonstrated potential advances toward resilient and decentralized spatial navigation algorithms.
The bottom-up mechanisms of spatial self-organization in such models can produce complex or chaotic behaviors.
For example, we previously introduced the \NS\ controller, in which agent-based interactions were modeled by analogy to neuronal network interactions, including spatial attractor dynamics and temporal phase-synchronization, that have been theorized to operate within the hippocampal place-cell circuits of navigating rodents.
While this complexity may drive emergent navigational capabilities, it also precludes linear analyses of stability, controllability, and performance that are used to study conventional swarming models.
Further, tuning dynamical controllers by hand or grid search is often inadequate due to the potential complexity of objectives, dimensionality of model dynamics and parameters, and computational costs of continuous-time simulation-based sampling.
Here, we present a generalizable framework for tuning dynamical controller models of autonomous multi-agent systems based on \ac*{bo}.
Our approach utilizes a task-dependent objective function to train \acp*{gp} as surrogate models to achieve adaptive and efficient exploration of a dynamical controller model's parameter space.
We demonstrate this approach by applying an objective function for \NS\ behaviors that cooperatively localize and capture spatially distributed rewards under time pressure.
We generalized task performance to new environments by combining objective scores for simulations in geometrically distinct arenas for each sample point.
To validate search performance, we compared high-dimensional clustering for high-~vs.~low-likelihood parameter points by visualizing surrogate-model trajectories in \ac*{umap} embeddings.
Our findings show that adaptive, sample-efficient evaluation of the self-organizing behavioral capacities of complex systems, including dynamical swarm controllers, can accelerate the translation of neuroscientific theory to applied domains.
\end{abstract}



\begin{keyword}
Bayesian Optimization \sep active learning \sep multi-agent swarming 
\sep phase synchronization \sep dynamical systems \sep spatial navigation
\end{keyword}

\end{frontmatter}

\section{Introduction}
\label{sec:introduction}

Roboticists have long turned to biological inspiration, including
swarming behaviors like flocking and schooling~\cite{Couz09}, to address
the problem of decentralized control and coordination of autonomous
multi-agent groups~\cite{beni2004swarm, cashin2004swarm, brambilla2013swarm,
bayindir2016review, hasselmann2018automatic, brown2018discovery}. However, the
general problem of how to analyze such swarming models has become relevant due
to the increasing prevalence of dynamical systems simulations in computational
studies of multi-agent interactions~\cite{coppola2018optimization}. The
field of \ac{ai} has shown impressive success by applying a minimal set of
brain-inspired concepts to connectionist learning models~\cite{lecun2015deep};
however, continuing to advance this biological inspiration, e.g., by integrating
temporal features of complex brain dynamics, faces both tremendous uncertainty
and potential reward~\cite{MonaRaja21}. Thus, progress on critical questions
necessary to advance multiple domains from robotics to \ac{ai} may depend on
methods for efficient computational characterization of complex dynamics in
models across multiple scales. More principled and interpretable methods are
needed that minimize the computational costs and automate the analysis of models
of nonlinear many-agent interactions.

\ac{bo} is a probabilistic modeling framework for adaptive, sample-efficient
optimization (e.g., expectation maximization) and evaluation (e.g., active
learning) of model parameter spaces. In this framework, model performance
is quantified by a task-dependent objective function and the sampling
trajectory is guided by an acquisition function that proposes the next
sample point based on a surrogate model. The surrogate model is typically
implemented as a \ac{gp} that spans some parameter subspace of interest
with multivariate normal distributions~\cite{rasmussen2003gaussian,
SnoeLaro12}, the means and variances of which are updated with each sample
evaluation to reflect the expected values and uncertainty, respectively,
of the underlying model's performance. Recent \ac{bo} studies have
demonstrated acceleration of hyperparameter tuning and optimization of
evolutionary algorithms, massively multi-modal functions, and other complex
models~\cite{roman2016bayesian, nguyen2019bayesian,roman2019bayesian} and
applications in robotic control~\cite{kieffer2018bayesian, rai2019using,
berkenkamp2021bayesian}. However, as parameter size grows, the computational
cost of the matrix inversions required to calculate updated \ac{gp} parameters
increases exponentially and eventually outweighs the gains in adaptive search
efficiency provided by computing the acquisition function over the surrogate
model to advance the sample trajectory~\cite{rasmussen2003gaussian}. This
limitation on model dimensionality does not in general prohibit analysis of
complex dynamics, particularly in systems of homogeneous particles, but it
would reasonably detract the feasibility of \ac{bo} for modeling systems with
nontrivial heterogeneity in agent/particle behaviors. Within that moderate limit
on model complexity, \ac{bo} may facilitate adaptive and efficient computational
exploration of dynamical parameter spaces, resulting in the illumination of the
reachable landscape of possible system behaviors.

The collective behavioral states of some swarming models, including
`swarmalators' based on oscillatory phase-coupled spatial
interactions~\cite{OKeeHong17, OKeeBett19}, is tractable to linear analysis of
stability, density, and clustering properties~\cite{IwasIida10, IwasTana10,
OKeeCero21} independent of computer simulation-based evaluation However,
for the potentially unbounded class of dynamical systems that preclude such
analysis due to nonlinearity, stochasticity, or other features, the time
and cost budgets to conduct system identification, active learning, and/or
optimization based on computational simulation-based samples are limiting
factors in translation to engineered designs in applied domains. Further,
emergent collective behaviors such as swarming outstrip the limitations 
of conventional agent-based task learning approaches including optimal 
policy selection under \ac{rl}. Thus, sample-efficient strategies are 
required. Our previous paper~\cite{monaco2020cognitive} introduced the \NS\
framework to demonstrate high-level emergent navigational behavior in a 
brain-inspired multi-agent metacontroller model. The number of dynamical
parameters in this model provided sufficient complexity to preclude analytic 
solutions and hand-tuning to assess and optimize global behaviors of the 
system. \NS\ behaviors included decentralized swarming and distributed 
goal-approach navigation in simulated environmental arenas with irregular,
complex, and/or fragmented geometry. However, due to the computational 
infeasibility of typical numerical optimization methods described above, these
findings~\cite{monaco2020cognitive} were based on extensive manual parameter
tuning, search, and exploration of qualitative \NS\ behaviors. Here we show how
neurodynamical controller models with emergent properties can be characterized
and optimized using \ac{bo} with \ac{gp} surrogate models.

\section{Theoretical Background of Neuroswarms}
\label{sec:theoretical_background}

In computational neuroscience, previously unachievable insights into theoretical
mechanisms may be enabled by defining objective functions that efficiently
discover and characterize the complex behaviors that emerge in high-dimensional
neural systems. We demonstrate a method for efficiently exploring the parameter
space of complex swarming behaviors as guided by an objective function that
measures cooperative foraging.

Distributed control and cooperation of multi-agent systems has been
a challenging problem within the fields of Robotics and Artificial
Intelligence, encouraging exploration of neuroscience-inspired paradigms.
Based on recent discoveries in neuroscience~\cite{monaco2019spatial,
monaco2019cognitive},~\cite{monaco2020cognitive} developed a novel theory for
the control of self-organized multi-agent systems within a two-dimensional
environment simulating responses to homogeneous data streams based on
visual cues. This framework, coined \NS, was intended for autonomous swarm
control driven by synaptic learning rules that treat multi-agent groups
analogically to neural network models of spatial cognitive circuits in
rodents. This work exemplifies how theoretical neuroscience, including
self-organizing attractor~\cite{zhang1996representation} and synchronization
dynamics~\cite{monaco2019spatial, monaco2011sensory}, can be leveraged to
emergently control complex systems of spatially distributed and decentralized
artificial agents.~\NS\ employs phase-based organization inspired by
the oscillatory state of hippocampal neurons with respect to the theta
rhythm (5–12 Hz) that is continuously active when rats engage in spatial
tasks~\cite{buzsaki2005theta}. Generalized phase codes such as the allocentric
modulation of phaser cells~\cite{monaco2019spatial} or the idiothetic modulation
of networks of velocity-controlled oscillators~\cite{monaco2011sensory,
blair2014oscillatory} were hypothesized to provide a minimal, resilient, and
efficient basis for a distributed tractable computation framework.

The key realization is that each agent ($x_i$) can be represented with a spatial
neuron (e.g., place cell or phaser cell that code for location) within the
cognitive map of the rodent’s brain. The strength of the synaptic connections
($W^{\prime}_{ij}$) between spatial cells determines the preferred distances
between neural agents ($D^{\prime}_{ij}$), expressed as:
\begin{equation}
    D^{\prime}_{ij} = -\alpha \log W^{\prime}_{ij}.
\end{equation}
Therefore, changes in agent positions during swarming is directly represented as
changes in synaptic weights between agents ($\alpha$ is a spatial constant). For
$N_s$ agents, new increments of spatial location of agents ($d_{xi}$) defined
as:
\begin{equation}
    d_{xi} = \frac{1}{2 \sum_{j} V_{ij}} \sum^{N_s}_{j=1} V_{ij} \
      \left ( D^{\prime}_{ij} - D_{ij} \right) \frac{x_j - x_i}{|x_j - x_i|},
\end{equation}
can be translated to corresponding new positions in the environment for
mutually visible agent-pairs ($V_{ij}$), in relation to the difference
between preferred and actual ($D_{ij}$) neural agent distances. With this
insight,~\cite{monaco2020cognitive} extended a conventional swarming formalism
with the addition of Hebbian learning to derive \NS, demonstrating for the first
time that swarming is network-based learning.

\section{Methods}
\label{sec:methods}

The complete \NS model is described in detail in~\cite{monaco2020cognitive}
given by equations 3-24. This model contains of 23 parameters in total, most of
which can be held constant across multiple environments. The 9 ($N_p$) tunable
parameters are described in Table~\ref{tbl:tunable_parameters} (and additional
constants listed in Table~\ref{tbl:constants} in the Appendix), which require
delicate tuning to properly balance the swarming and reward capture mechanics of
\NS. There is no closed-form solution to this dynamical system and, conversely,
there are too many parameters to efficiently search using naive computational
methods, such as grid search. Thus we investigated several adaptive optimization
methods under the umbrella of \ac{bo}.

\begin{table}[htb!]
\scriptsize
    \begin{center}
        \ra{1.3}
        \begin{tabular}{@{} lrl @{}}
        \toprule
            Parameter & Range & Description \\
        \midrule
            $\sigma$ & $[10^{-3}, 4]$ & Swarm interaction spatial scale \\
            $\eta_s$ & $[10^{-3}, 4]$ & Swarm connections learning rate \\
            $\eta_r$ & $[10^{-3}, 4]$ & Reward connections learning rate \\
            $\kappa$ & $[10^{-3}, 4]$ & Reward interaction spatial scale \\
            $\omega_0$ & $[0, 1]$ & Baseline oscillatory frequency \\
            $\omega_I$ & $[0, 1]$ & Max increase in oscillatory frequency \\
            $\tau_q$ & $[0, 1]$ & Swarming inputs time constant \\
            $\tau_r$ & $[0, 1]$ & Reward inputs time constant \\
            $\tau_c$ & $[0, 1]$ & Sensory-cue inputs time constant \\
        \bottomrule
        \end{tabular}
    \end{center}
\caption{Tunable Parameters that govern the dynamical behavior of a \NS\ model.}
\label{tbl:tunable_parameters}
\end{table}

\subsection{Bayesian Optimization}

\begin{figure}[tb!]
    \centering
    \begin{subfigure}{0.7\textwidth}
        \centering
        \caption{\acl{gp} Surrogate Model Training}
        \includegraphics[width=\linewidth,trim=0cm 2cm 0cm 0.2cm,clip]{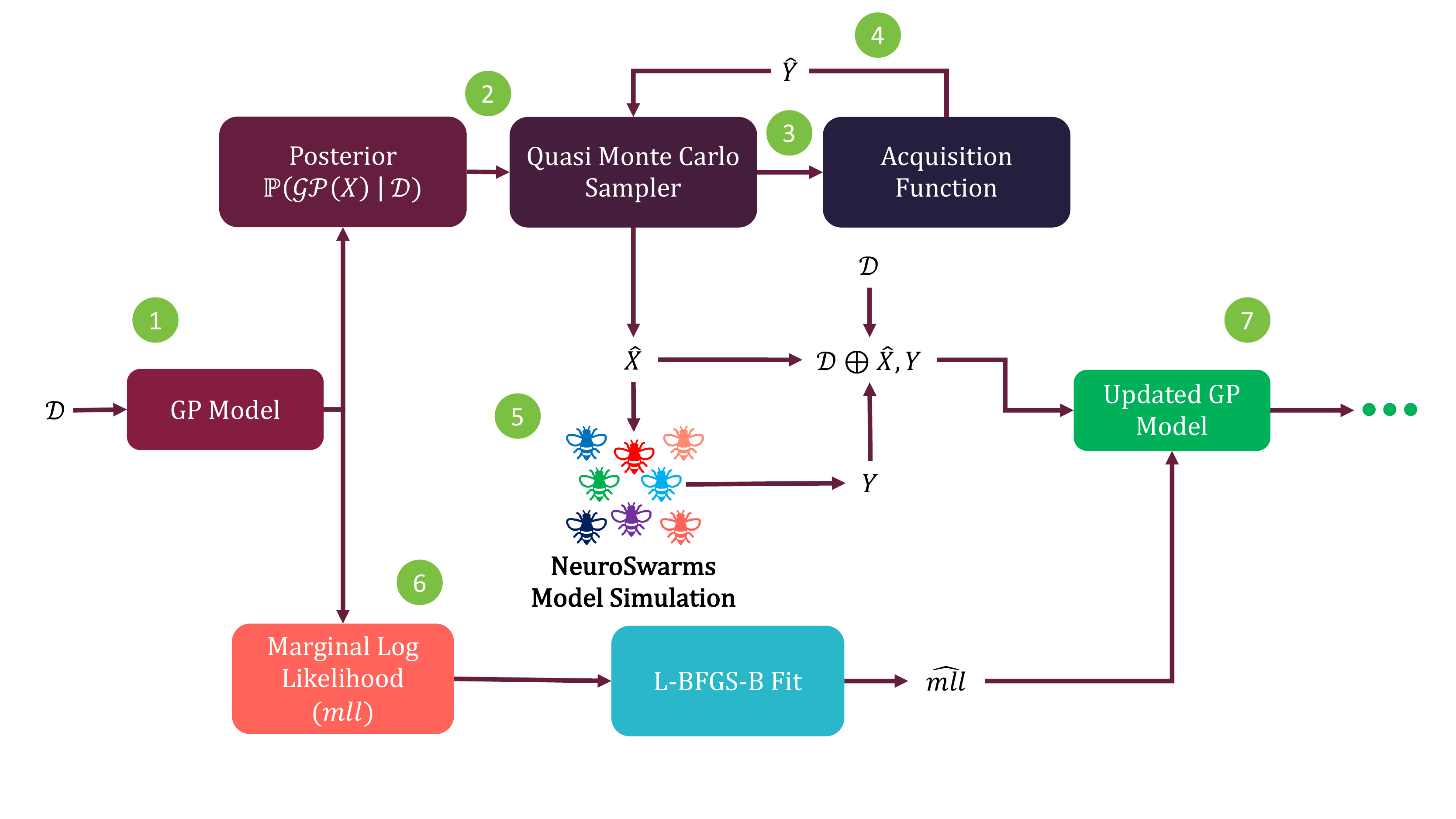}
        \label{fig:block_diagram}
    \end{subfigure}
    \begin{subfigure}{0.55\textwidth}
        \centering
        \caption{\NS\ Simulation-based Sampling (Step 5)}
        \includegraphics[width=\linewidth,trim=4.5cm 0.6cm 3.5cm 0cm,clip]{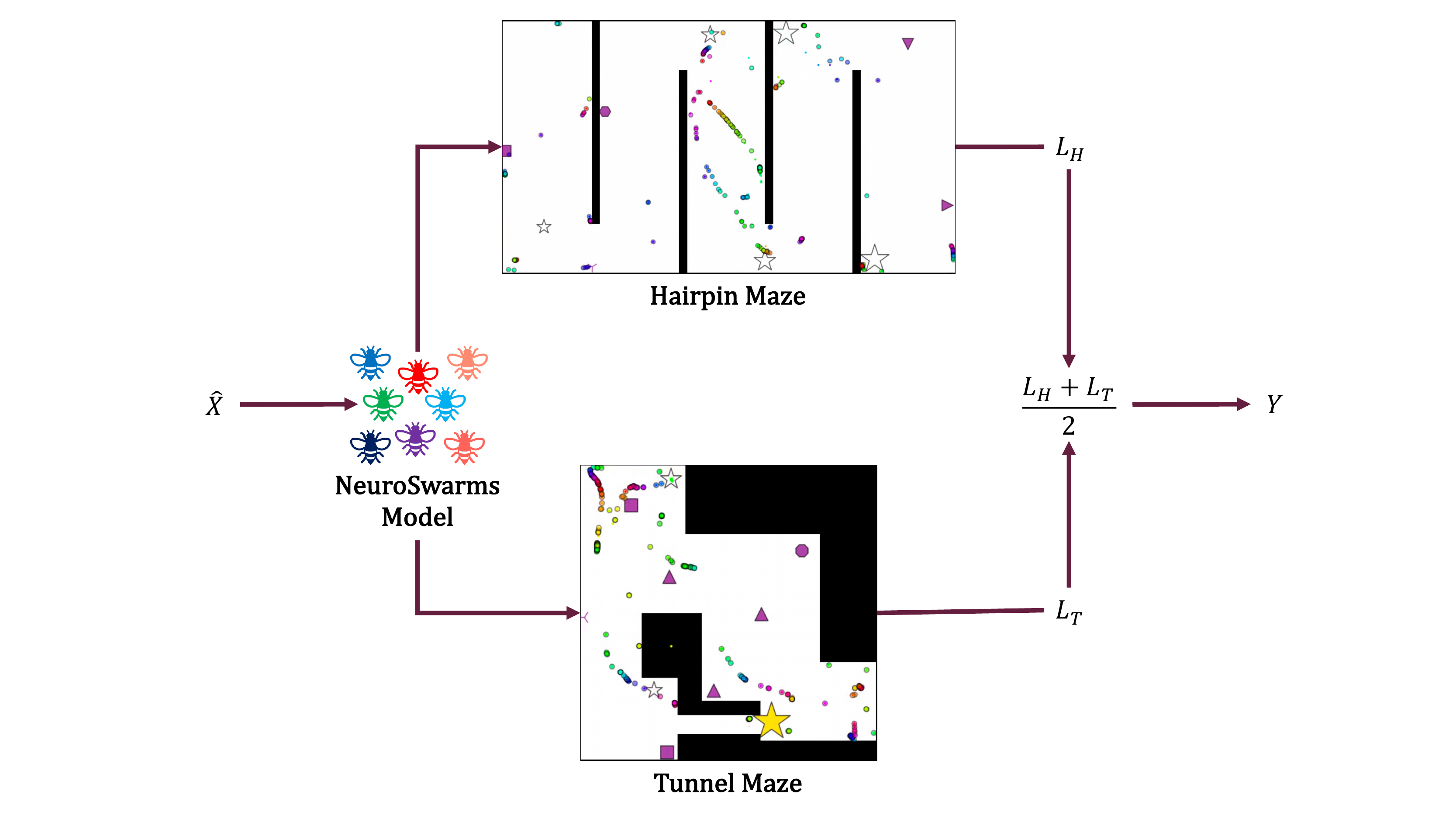}
        \label{fig:NS_block_diagram}
    \end{subfigure}
    \caption{%
(a.) Computation flow for optimization and simulation-based sampling. Step 1:
The posterior distribution is generated from the \ac{gp} model based on the
preliminary training data ($\mathcal{D}$) and the set of candidate parameters.
Step 2: The quasi-\ac{mc} sampler of the acquisition function uses the posterior
distribution to select future candidate parameters ($\hat{X}$) based on the
(Steps 3 and 4) acquisition function's calculated average utility ($Y$). Step
5: The \NS\ model simulates using $\hat{X}$ which results in a corresponding
objective performance value ($Y$). Step 6: The initial \ac{gp} model's marginal
log likelihood ($mll$) is calculated and then used to optimize the \ac{gp} model
using the limited memory Broyden–Fletcher–Goldfarb–Shannon algorithm
with simple bounds (L-BFGS-B~\cite{zhu1997algorithm}). Step 7: The resulting
components are then used to construct a new \ac{gp} model and calculate an
updated marginal log likelihood ($mll$). (b.) Flow diagram of simulation-based
sampling. During a simulation, the \NS\ model executes a play-through on the
Hairpin and Tunnel Mazes and then averages the losses ($L_H$ and $L_T$) across
each run to calculate $Y$.
\acresetall
}
\end{figure}

\ac{bo} is a technique that allows for the construction and sequential
optimization of surrogate models to represent the objective performance of more
complex models~\cite{o1978curve, jones1998efficient, osborne2010bayesian}.
Optimizing surrogate models can be beneficial in scenarios where executing
a complex model is slow and/or computationally expensive. These surrogate
models can be deployed to predict the performance of the underlying
complex model at untested parameter values without actually running it.
Figure~\ref{fig:block_diagram} illustrates the \ac{bo} process we use to
optimize the \NS\ model.

Specifically, in our work we explore a category of \ac{bo} surrogate models
called \ac{gp} models~\cite{rasmussen2003gaussian, williams1998prediction,
mackay1997gaussian}. \ac{gp} models are probabilistic parametric
models that learn to estimate objective performance from limited
observations~\cite{SnoeLaro12}. \ac{gp} models iteratively learn a probabilistic
mapping from $n$ input parameter points $x_i \in \mathbb{R}^{N_p}$ to
the desired output objective function value $y_i \in \mathbb{R}^{N_y}$
($N_y = 1$ in the case of our scalar objective) expressed as $p(y_i
| x_i)$~\cite{krauth2016autogp, balandat2020botorch}. \ac{gp} models
estimate this mapping by assuming the complex model $f_{\mathrm{true}}$ is
distributed according to a \ac{gp}: \begin{equation} f_{\mathrm{true}} \sim
\mathcal{GP}_{\mu, k}( \mathbf{X}), \end{equation} where $\mu$ and $k$ are
the mean and covariance kernel on $x_i \in \mathbf{X}$, and $\mathbf{X}$ is
set of training points used to seed the GP model. The posterior distribution
$P(\mathcal{GP}(X) | \mathcal{D})$ of candidate set $X = \{\xi_1, \cdots,
\xi_q\}$, where $q$ is the number of candidate points, conditioned on
data $\mathcal{D} = \{(x_i, y_i)\}^n_{i=1}$ is a multivariate normal
$\mathcal{N}(\mu(X),k(X))$.

The other key component of the \ac{bo} process is an acquisition function,
which is used to navigate the underlying complex model's parameter space.
Acquisition functions define a strategy to manage the trade-off between
exploring the parameter space and exploiting promising regions of the parameter
space that yielded improvement in previous samples~\cite{shahriari2015taking}.
The acquisition function's parameters $\Phi$ are optimized by evaluating it on
$P(\mathcal{GP}(X) | \mathcal{D})$. Formally, an arbitrary acquisition function
($\alpha$) can be expressed as follows:
\begin{equation}
\alpha(X; \Phi, \mathcal{D}) = \mathbb{E}[\mathtt{a}(g(\mathcal{GP}(X)), \Phi) | \mathcal{D}],
\label{eq:analytical_acq_func}
\end{equation}
where $\mathtt{a}$ is the utility function and $g$ is a composite objective
function~\cite{wilson2018maximizing}. However, analytical expressions
are not always available for arbitrary $\mathtt{a}$ or $g$, which is why
\ac{mc} integration is used to approximate the expectation through sampling
$P(\mathcal{GP}(X) | \mathcal{D})$, resulting in:
\begin{equation}
    \hat{\alpha}_N(X; \Phi, \mathcal{D}) = \frac{1}{N}\sum_{i=1}^{N} \mathtt{a}(g(\mathcal{E}^i_{\mathcal{D}}(X)), \Phi).
\end{equation}

The approximated acquisition function $\hat{\alpha}_N$ uses $N$ number of MC
samples used to approximate $\mathcal{E}^i_{\mathcal{D}} \sim P(\mathcal{GP}(X)
| \mathcal{D})$. We experiment with a pair of MC acquisition functions
q-Expected Improvement (qEI) and Noisy q-Expected Improvement (qNoisyEI),
compared against random search. qEI can be expressed as:
\begin{equation}
    \text{qEI}(X) \approx \frac{1}{N} \sum^{N}_{i=1} \max_{j=1,\cdots,q} \{\max( \mathcal{E}^i_j - Y^*, 0)\},
\end{equation}
where $q$ batches are sampled from the joint posterior, whereby the improvement
over the current best observed objective function value ($Y^*$), assuming
noiseless observations, is calculated for each sample, and the maximum is
taken for each batch and averaged across the number of MC samples. As in the
acquisition approximation, $\hat{\alpha}_N$, qEI approximates the posterior with
$\mathcal{E}^i \sim P(\mathcal{GP}(X) | \mathcal{D})$.

Similarly, qNoisyEI~\cite{letham2019constrained} operates on the same principle
of averaging the maximum improvement for $q$ batches of samples, except it
does not assume noiseless observations such as $Y^*$ being the best objective
function value. Instead,
\begin{equation}
    \text{qNoisyEI}(X; \mathcal{D}) \approx \frac{1}{N} \sum^{N}_{i=1} \max_{j=1,\cdots,q} \{\max \mathcal{E}^i_j - \max \mathcal{E}_{obs}, 0\},
\end{equation}
where $(\mathcal{E}, \mathcal{E}_{obs}) \sim \mathcal{GP}((X, X_{obs}))$, $\mathcal{E}_{obs}$ is an approximation of the posterior distribution of previously observed parameters $X_{obs}$.

\subsection{Objective Function}

We introduce an objective function to evaluate the performance of a multi-agent
model (e.g. \NS) in time-optimal cooperative reward capture. The objective
function quantifies how quickly the swarm of agents can forage and collectively
capture all the rewards in a given environment. Let $N^{\text{cap}}(t)$ be
the number of cooperatively captured rewards where a reward is captured if at
any time-step $t$ at least $N_s / N_r$ number of agents were simultaneously
colocated within a defined radius from the reward. This objective function can
be expressed as a loss,
\begin{equation}
    L = -t / \left(N_t N^{\text{cap}}(t) + 1 \right),
    \label{eq:NS_obj_func}
\end{equation}
which is updated at each time-step, where $N_t$ is the total number of time
steps, until all $N_r$ rewards have been captured. The swarm is encouraged to
complete the task quickly since $t$ grows for each time-step needed to capture
all the rewards. If the swarm is not able to capture all the rewards in the
environment, $t$ will be set to the maximum number of time-steps allowed for the
simulation $N_t$ and the loss will only reduce for each $N^{\text{cap}}(t)$.
Objective values range from $[-1, 0]$, with the theoretical best value at zero.

When evaluating \NS\ simulation performance for a given set of parameters,
generalizable performance is also taken into account. Each simulation
constitutes a play-through of the Hairpin and Tunnel mazes (see
Figure~\ref{fig:NS_block_diagram}) where the objective performance is $L_H$
and $L_T$, respectively. Thus, the overall performance, $Y$, of candidate set
parameters $\hat{X}$ is calculated as the average:
\begin{equation}
    Y = \frac{L_H + L_T}{2}.
\end{equation}

\subsection{\ac{gp} Surrogate Training}

We use the \ac{bo} software framework BoTorch~\cite{balandat2020botorch}
to implement the overall \ac{bo} process described in
Figure~\ref{fig:block_diagram}. The training process starts by taking an initial
set of training examples $D$ is used to initialize the GP model. The resulting
posterior distribution $P(\mathcal{GP}(X) | \mathcal{D})$ is then sampled using
a batched quasi-MC sampler, where an acquisition function determines the set
of future candidate \NS\ parameters $\hat{X}$ based on the perceived utility
$\hat{Y}$ for each set of parameters. The sampled parameters are bounded by the
range limits outlined in Table~\ref{tbl:tunable_parameters}. Next, the set of
future candidate \NS\ parameters undergo a \NS\ model simulation which results
in a \NS\ performance evaluation $Y$. Afterwards, the pair $\hat{X}$ and {$Y$}
are incorporated into $D$ which will be used to initialize the \ac{gp} model in
the next optimization step. The \ac{gp} model is optimized by first computing
its marginal log likelihood $mll$-acting as a loss function- applied to $X$ and
then fitting the hyperparameters of the \ac{gp} model using the limited memory
Broyden–Fletcher–Goldfarb–Shannon algorithm. The fitting process also
results in an updated marginal log likelihood $\hat{mll}$ used for the next
optimization step. The whole optimization process is repeated for $N_T$ epochs,
where $N_T$ is determined by satisfactory convergence.

Training examples and \NS\ model simulations are formed by running simulations
according to Figure~\ref{fig:NS_block_diagram}, where a set of randomly chosen
parameters (in this case represented by $\hat{X}$) are used to initialize the
\NS\ model and a unique simulation is performed on the hairpin and tunnel mazes
separately. \NS\ model performance is evaluated for each environment according
to Equation~\ref{eq:NS_obj_func} to encourage cooperative foraging, which is
then averaged to produce the generalized foraging performance $Y$. Simulations
are conducted on two unique mazes to evaluate how well the $\hat{X}$ results in
\NS\ model performance that generalizes across environments.

\subsection{Convergence Metrics}

In order to understand when \ac{gp} model optimization has converged to a
satisfactory conclusion we use two convergence metrics. One evaluation criteria
would be the maximum posterior variance for the $M$ epoch compared with all
previous epochs, which can be expressed as follows:
\begin{equation}
    \Var_{\max}(x_M, \mathcal{D}_M) = \max_{i=1}^M \Var(\mathop{P} (\mathcal{GP}(x_i) | \mathcal{D}_i)).
\end{equation}

This stopping criteria suggests that when the \ac{gp} model's posterior variance
is no longer increasing when compared to previous training iterations, then
training should stop because the \ac{gp} model is having minimal change which is
undesirable. Dissimilarity is another metric that couples well with the maximum
posterior variance because it measures the lack of change in the selection
of future candidate parameters $x_M \in \R^{M \times N_p}$ when compared to
previously simulated parameters $x_i$. Dissimilarity is defined as:
\begin{equation}
    D(X_M) = \min_{i=1}^M 1 - \frac{x_i \cdot x_{M}}{\max(\norm{x_i}_2 \cdot \norm{x_{M}}_2, \epsilon)}.
\end{equation}

When a given \ac{gp} model's dissimilarity approaches zero the set of parameters
from a given epoch is most similar to the set of parameters selected by the
acquisition function on the last epoch of training. In order to prevent division
by zero, we choose to take the maximum of the product of norms and a constant
$\epsilon=10^{-6}$. Overall, the two convergence metrics evaluate the change
of the \ac{gp} model in terms of its posterior variance or MC sampled $X_M$,
which correspond to improvement in the descriptive capability of the \ac{gp}
model's posterior. Moreover, a descriptive \ac{gp} model posterior is essential
to understanding the performance of \NS\ parameters in parameter space.

\subsection{Visualizing The Parameter Space}

Estimating the performance of \NS\ on a set of parameters via a \ac{gp} model
is useful, but understanding which parameters to choose is equally important.
Providing a method of visualizing performance across the parameter space
facilitates understanding the impact of regions of the parameter space on
performance. A visualization tool can allow for the discovery of pockets of
high performing parameter sets that solve the foraging problem in unique
ways. Since the multivariate norms are entangled in dependencies between
all $N_p$ parameters, we turn to UMAP~\cite{mcinnes2018umap} to embed the
high dimensional parameter space manifold into a 2D representation. The low
dimensional representation is trained to spatially cluster the parameter space
such that $N_p$ size sets of parameters are grouped together if they are similar
numerical values, and are pushed apart if they are numerically distant from each
other. This results in a simple visual representation of $X$ where each point
can be colored by the corresponding $Y_i$ or any of the corresponding parameter
values associated with said point. We use this visual tool to qualitatively
evaluate the performance of the \ac{gp} model in approximating the \NS\ model's
performance by selecting a point in UMAP space that has high performance (large
$y_i$) and matching that point to each of the corresponding parameter values
from the parameter set ($x_i$).

\section{Results and Discussion}
\label{sec:experiment_results}

We present experiments illustrating the use of Bayesian Optimization
to tune the parameters of a neuroscience-inspired swarming model,
\NS~\cite{monaco2020cognitive}, tasked with cooperatively capturing multiple
rewards in a maze. We train and optimize surrogate \ac{gp} models to explore
and characterize the parameter space of the swarming model using a series
of acquisition functions. Then we demonstrate how non-linear dimensionality
reduction via \ac{umap}~\cite{mcinnes2018umap} can be used to visualize the
parameter space and aid in identifying and selecting a set of parameters that
are likely to result in strong model performance. Last, we qualitatively
evaluate the performance of selected parameters at generalizing across two
distinct environments.

\subsection{Gaussian Process Model Training}

\begin{figure}[tb!]
    \centering
    \begin{subfigure}{0.49\textwidth}
        \centering
        \caption{Dissimilarity}
        \includegraphics[width=\linewidth,trim=0cm 0cm 0cm 0.5cm,clip]{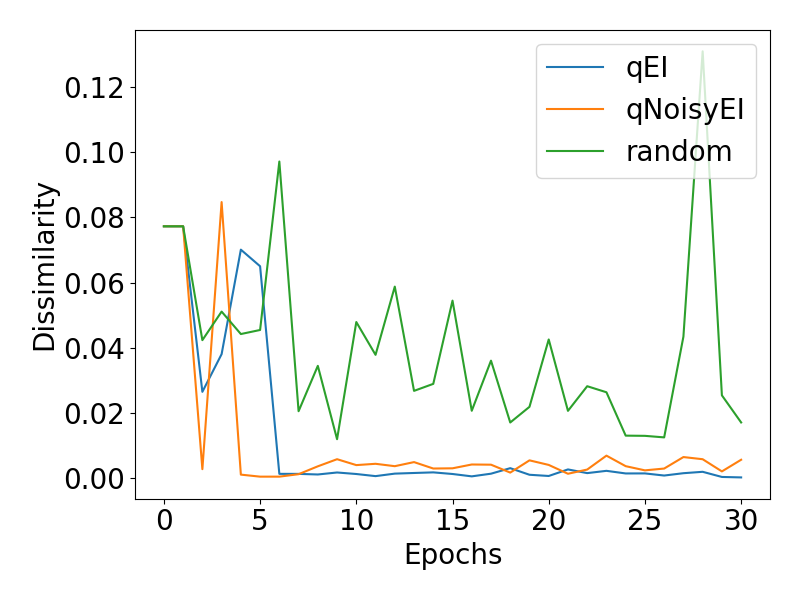}
        \label{fig:dissimilarity}
    \end{subfigure}
    \begin{subfigure}{0.49\textwidth}
        \centering
        \caption{\ac{gp} Posterior Variance}
        \includegraphics[width=\linewidth,trim=0cm 0cm 0cm 0.5cm,clip]{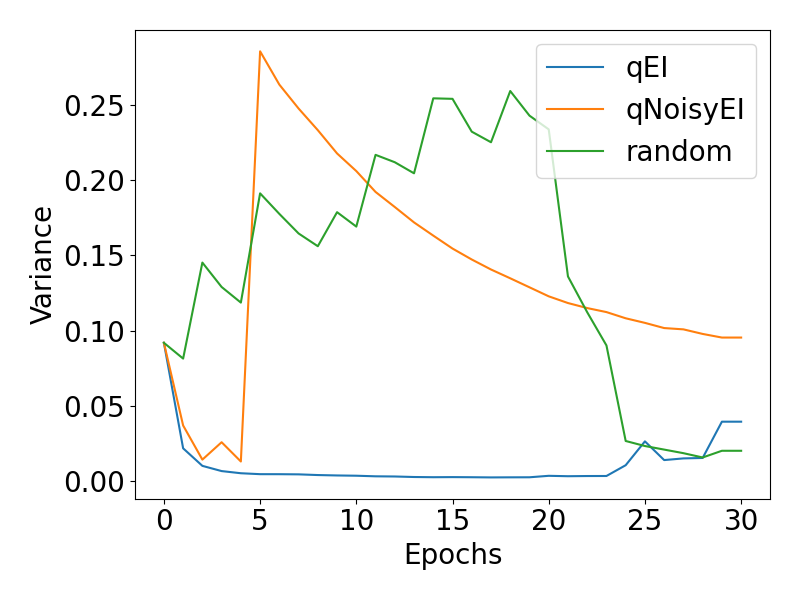}
        \label{fig:variance}
    \end{subfigure}
    \vspace{-0.8cm}
    \caption{Acquisition function parameters dissimilarity per epoch compared
with final selected parameters.}
\end{figure}

Manually tuning swarming models, such as \NS, can be challenging as
small variations in the $N_p$ interdependent parameters ($N_p = 9$, see
Table~\ref{tbl:tunable_parameters}) can dramatically impact model behavior. An
optimal set of parameters that allows the \NS\ model to perform successfully
on two distinct environments may not be limited to a single set of parameters
or may not exist, thus we propose exploring the parameter space in a sample
efficient manner using a \ac{gp} surrogate model. We utilize a \ac{mc}-based
acquisition function to sample the parameter space and optimize the \ac{gp}
model's predictive performance compared with actual \NS\ simulation results.

An overview of the \ac{gp} model training process is depicted in
Figure~\ref{fig:block_diagram}. We optimize \ac{gp} models on two environments:
a hairpin and a tunnel maze (see Figure~\ref{fig:NS_block_diagram}) in order to
find dynamical regimes with efficient foraging dynamics.

We started the training process of four \ac{gp} models with an initial set of
24 sets of randomly selected parameters and corresponding simulation results
($\mathcal{D}$). Each of the three \ac{gp} models had one of four acquisition
functions that was used to explore the parameter space: \acf{ei}, \acf{nei},
and random sampling. \ac{gp} modeling and training was implemented using
the BoTorch~\cite{balandat2020botorch} framework. Each \ac{gp} model with a
corresponding acquisition function was optimized with $N=512$ \ac{mc} samples
over $N_{T} = 30$ epochs. All the acquisition functions, except for random,
were verified as having converged by the end of the training process using the
dissimilarity and posterior variance metrics. Figure~\ref{fig:dissimilarity}
illustrates that the non-random acquisition function \ac{gp} models were able
to reach near zero dissimilarity during training. Similarly, the posterior
distribution's maximum variance for each of the \ac{gp} models had reached
convergence by the end of the training duration. We stopped training once
convergence had been reached to avoid incurring unnecessary simulation
computational costs.

\begin{figure}[tb!]
    \centering
    \begin{subfigure}{0.49\textwidth}
        \centering
        \caption{Observed objective Function Histogram}
        \includegraphics[width=\linewidth,]{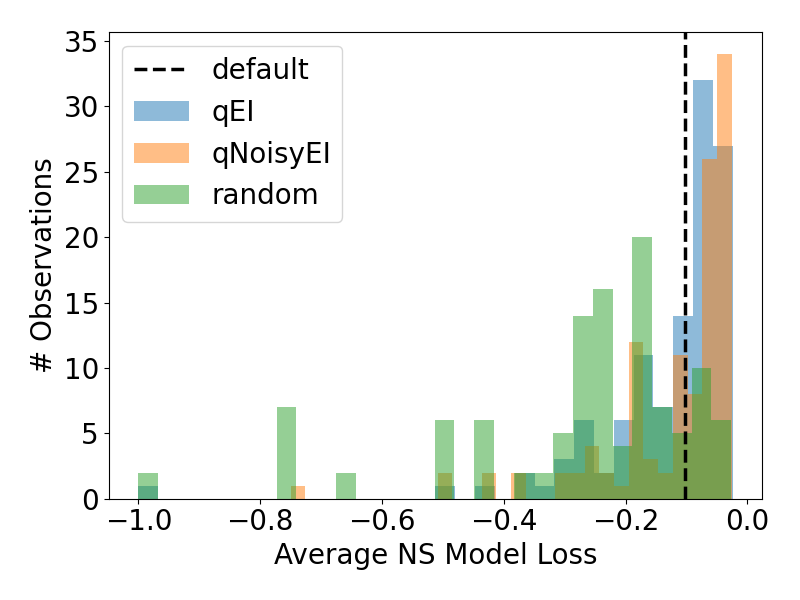}
        \label{fig:obs_histogram}
    \end{subfigure}
    \begin{subfigure}{0.49\textwidth}
        \vspace{-0.2cm}
        \centering
        \caption{Best Observed Value}
        \includegraphics[width=\linewidth,trim=1cm 7.1cm 1cm 7cm,clip]{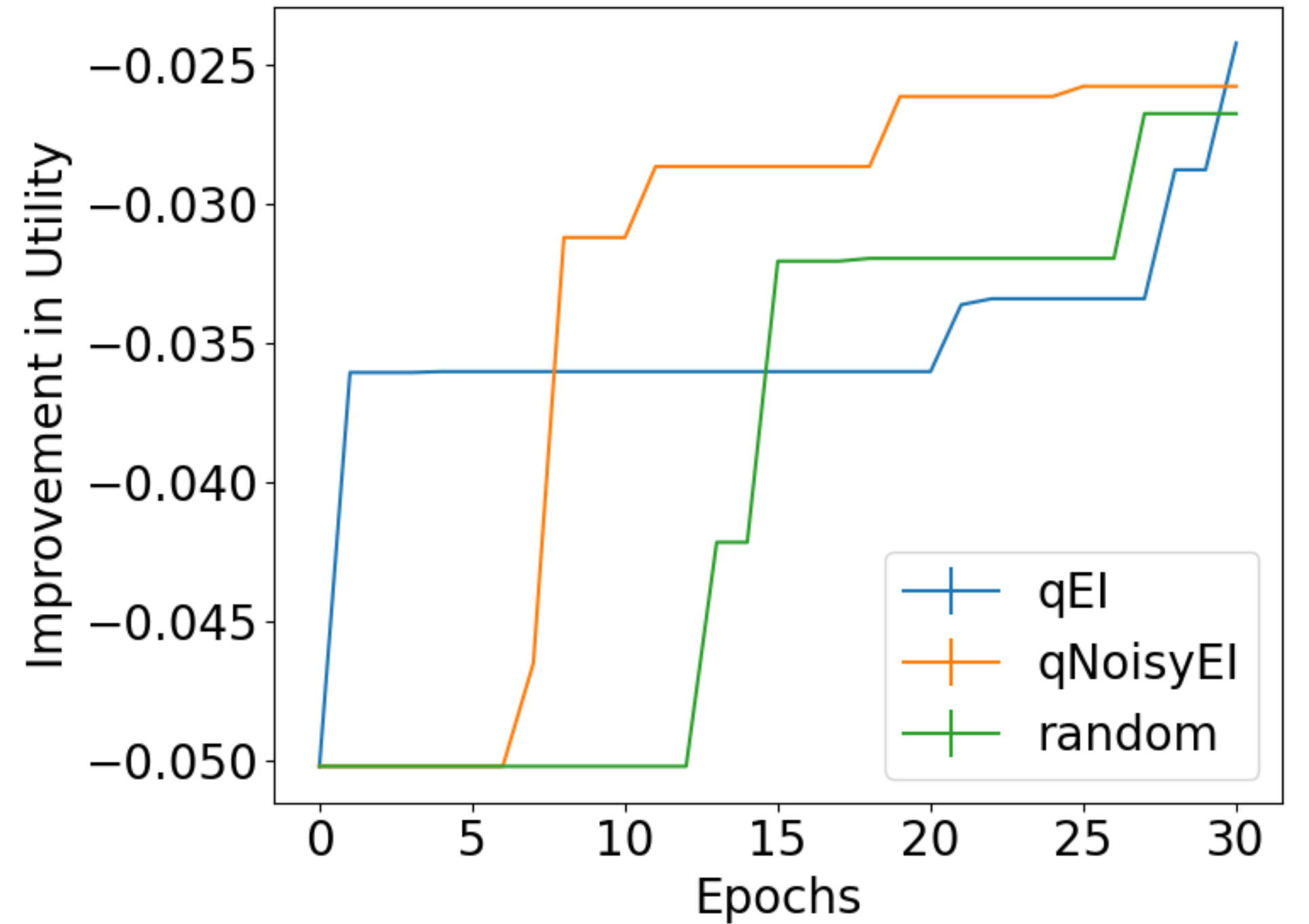}
        \label{fig:best_obs_val}
    \end{subfigure}
    \vspace{-0.6cm}
    \caption{\ac{gp} models with corresponding acquisition functions model
performances. Closest to 0 are the best.}
\end{figure}

Next, we evaluated how effective each \ac{gp} model's acquisition function
was at finding regions of the parameter space that maximize the \NS\
objective function. The histogram in Figure~\ref{fig:obs_histogram} shows
that the \ac{ei} and \ac{nei} acquisition function-based \ac{gp} models
were able to locate the most parameter sets that resulted in a high
performance \NS\ simulations (closest to 0 are best). For comparison,
random parameter selection and the default set of parameters (indicated
as default in Figure~\ref{fig:obs_histogram}) from Monaco et~al.
(2020)~\cite{monaco2020cognitive} were both out-performed by the \ac{ei}
and \ac{nei} acquisition functions-based \ac{gp} models. The \ac{gp} models
were optimized over fewer parameters with fewer degrees of freedom and still
outperformed the default manually tuned parameters. While our objective was
to explore the parameter space and not necessarily to find the optimal set of
parameters, \ac{ei} and \ac{nei} acquisition functions operate on expected
improvement in utility which resulted in the most utility improvement when
exploring the parameter space over the training process, as depicted in
Figure~\ref{fig:best_obs_val}.

\subsection{Qualitative Evaluation}
\label{ssec:qual_eval}

\begin{figure*}
    \centering
    \includegraphics[width=\linewidth,trim=11cm 3cm 9cm 2cm,clip]{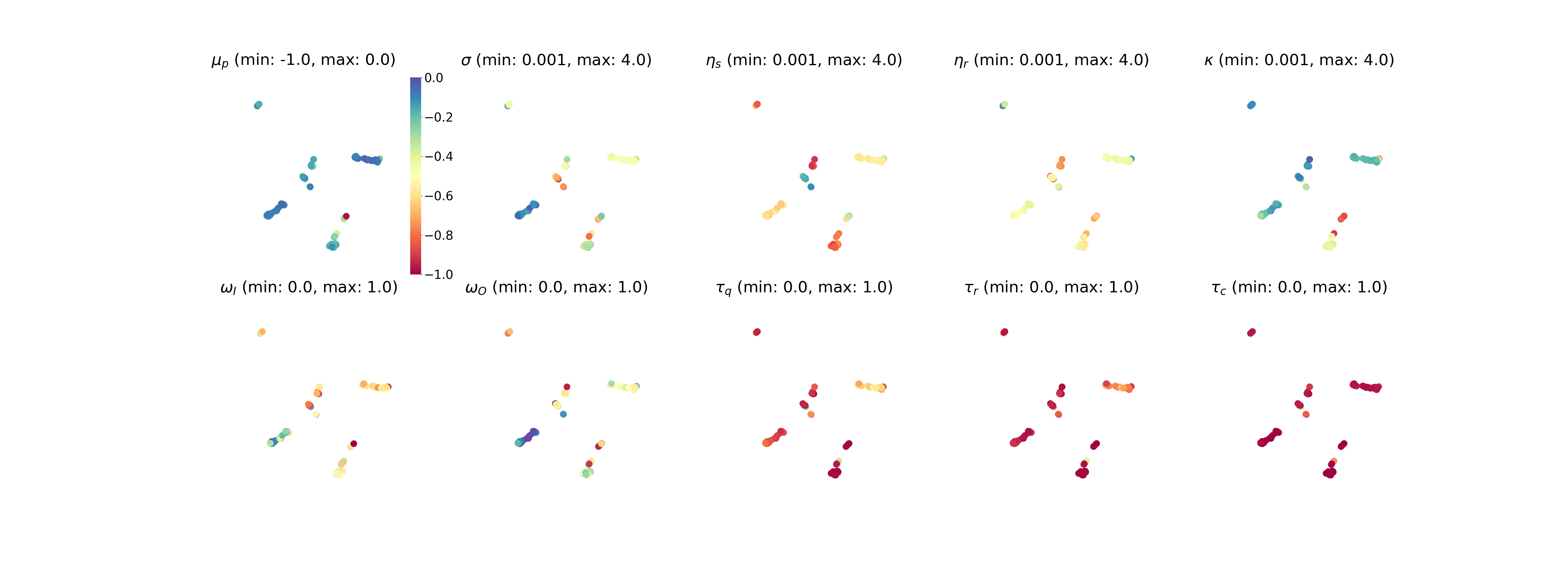}
    \caption{q-Expected Improvement (qEI) parameter space}
    \label{fig:qEI_params}
\end{figure*}

Utilizing the results of the Bayesian optimization process requires a visual
representation of the parameter space. Representing an $N_p$-dimensional
parameter space can be challenging, so we elected to use a dimensionality
reduction and visualization method called \ac{umap}~\cite{mcinnes2018umap}.
\ac{umap} allows the $N_p$-dimensional parameter space to be reduced to two
dimensions, which can then be easily visualized. As illustrated in the top left
plot in Figure~\ref{fig:qEI_params}, we were able to assign a color to each low
dimensional representation dot based on the corresponding utility value obtained
from the GP models' posterior distribution's mean value at said position in the
parameter space. Similarly, we generated plots for each of the parameters, where
points in parameter space were colored based on the corresponding value of said
parameter. As a result, we were able to create a visual representation where the
highest utility (posterior mean) set of points could be identified as a group,
and then individually using the $N_p$ other parallel plots.

Given that the \ac{gp} model corresponding to the \ac{ei} acquisition function
had the largest improvement in utility during the training process according to
Figure~\ref{fig:best_obs_val} and consistently identified high performance sets
of parameters (see Figure~\ref{fig:obs_histogram}), we used its depiction of the
parameter space for further analysis. Figure~\ref{fig:qEI_params} illustrates
that the \ac{ei} \ac{gp} model identified two clusters of sets of parameters
that have the highest utility. From the posterior mean plot, we selected one
point from the bottom left cluster and we matched it with the same corresponding
position in each of the individual parameter plots below. The color bar was
then used to identify the numerical value for all $N_p$ parameters, which were
then assigned to the \NS\ model for evaluation. In summary, instead of hand
tuning on each environment, the \ac{bo} process is an illustration of how we
simultaneously tuned all $N_p$ parameters for all environments.

\begin{figure}[tb!]
    \centering
    \begin{subfigure}{0.336\textwidth}
        \centering
        \caption{R5 Capture $t \leq 6.22$s}
        \vspace{-0.2cm}
        \includegraphics[width=\linewidth]{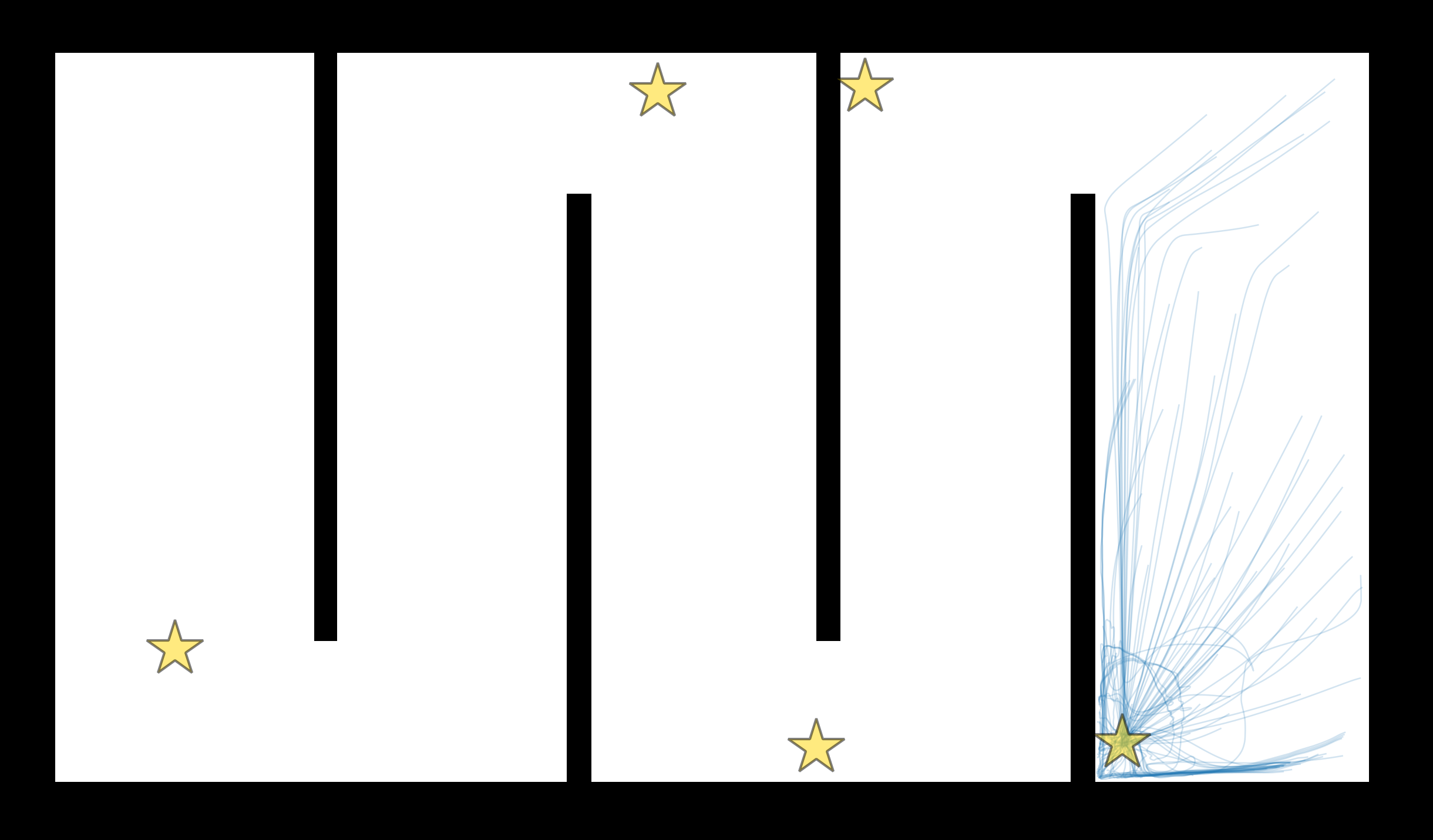}
        \label{fig:hairpin_traceplot_r5_capture}
    \end{subfigure}
    \begin{subfigure}{0.336\textwidth}
        \centering
        \caption{After R5 Capture $t > 6.22$s}
        \vspace{-0.2cm}
        \includegraphics[width=\linewidth]{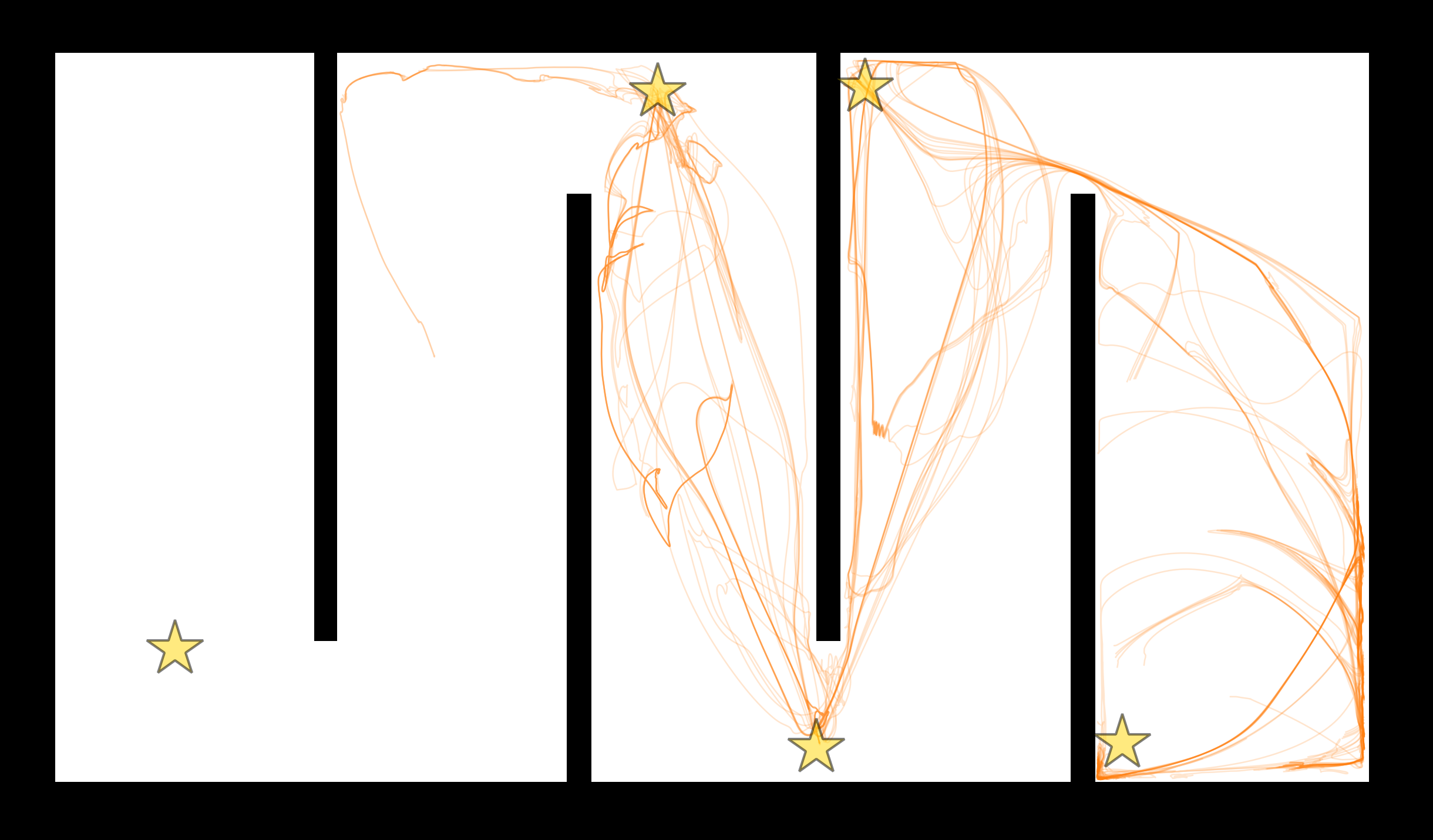}
        \label{fig:hairpin_traceplot_r5_after}
    \end{subfigure}
    \begin{subfigure}{0.336\textwidth}
        \centering
        \vspace{-0.2cm}
        \caption{R1 Capture $t \leq 9.53$s}
        \vspace{-0.2cm}
        \includegraphics[width=\linewidth]{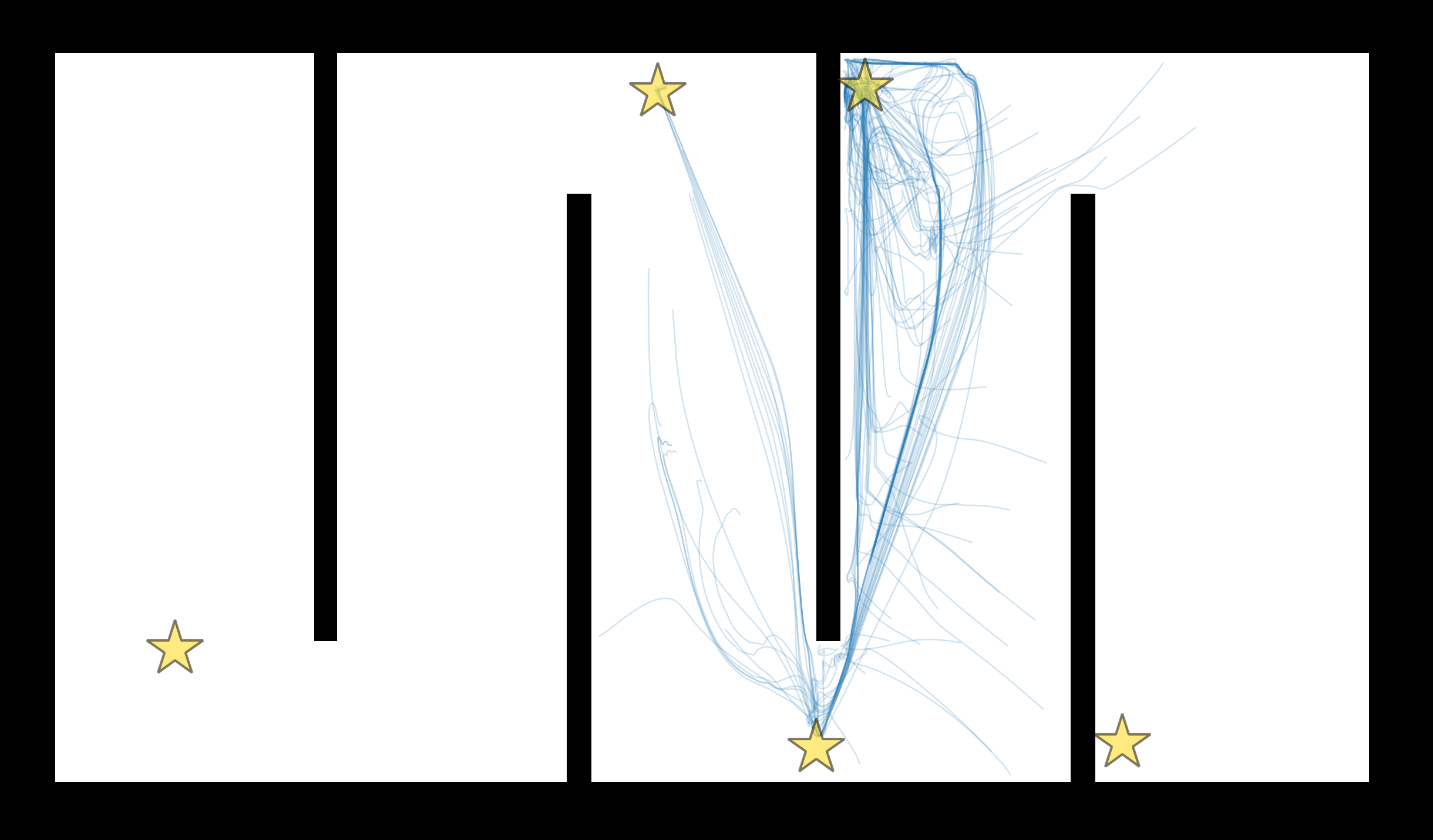}
        \label{fig:hairpin_traceplot_r1_capture}
    \end{subfigure}
    \begin{subfigure}{0.336\textwidth}
        \centering
        \vspace{-0.2cm}
        \caption{After R1 Capture $t > 9.53$s}
        \vspace{-0.2cm}
        \includegraphics[width=\linewidth]{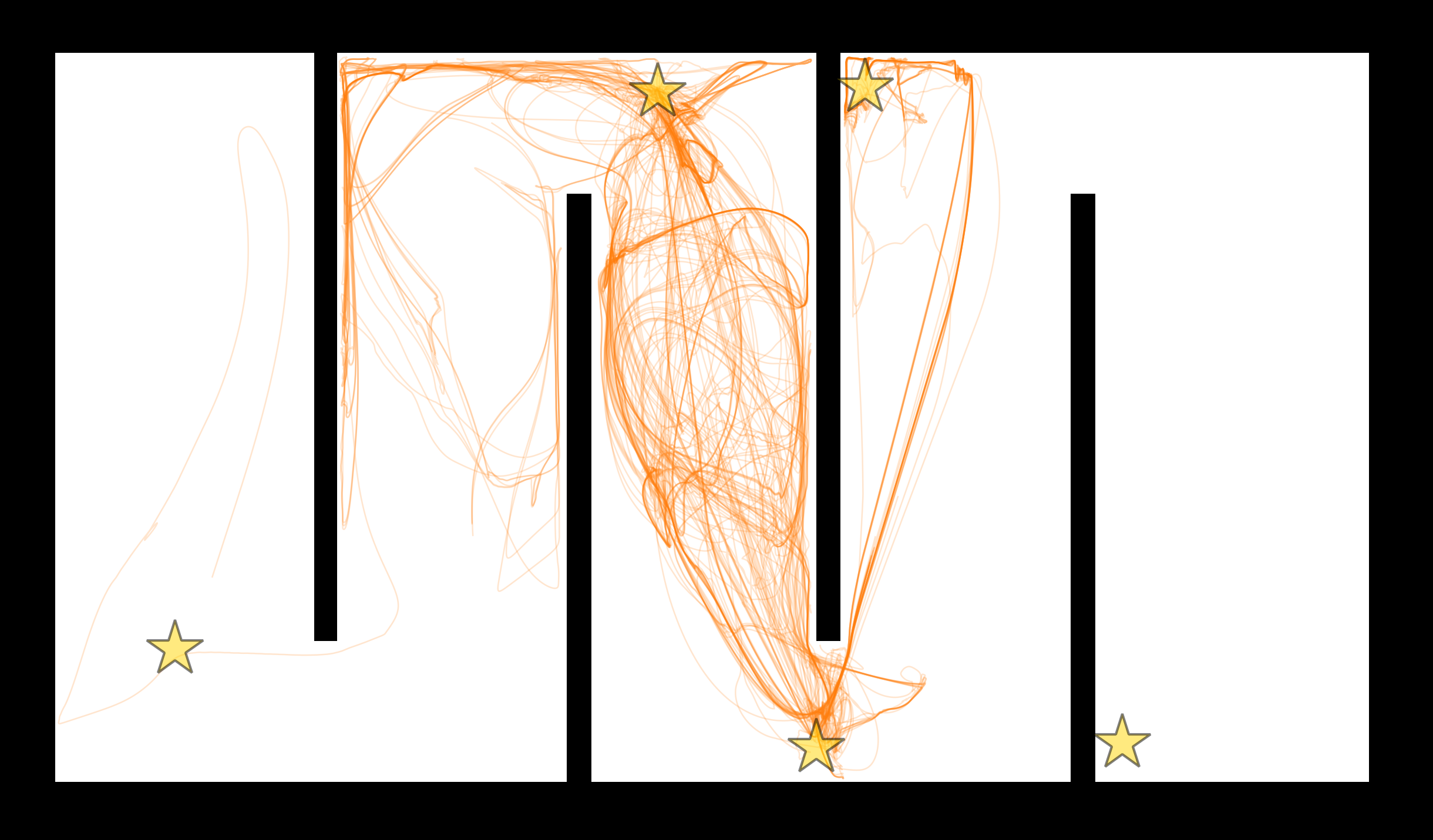}
        \label{fig:hairpin_traceplot_r1_after}
    \end{subfigure}
    \begin{subfigure}{0.336\textwidth}
        \centering
        \vspace{-0.2cm}
        \caption{R3 Capture $t \leq 11.86$s}
        \vspace{-0.2cm}
        \includegraphics[width=\linewidth]{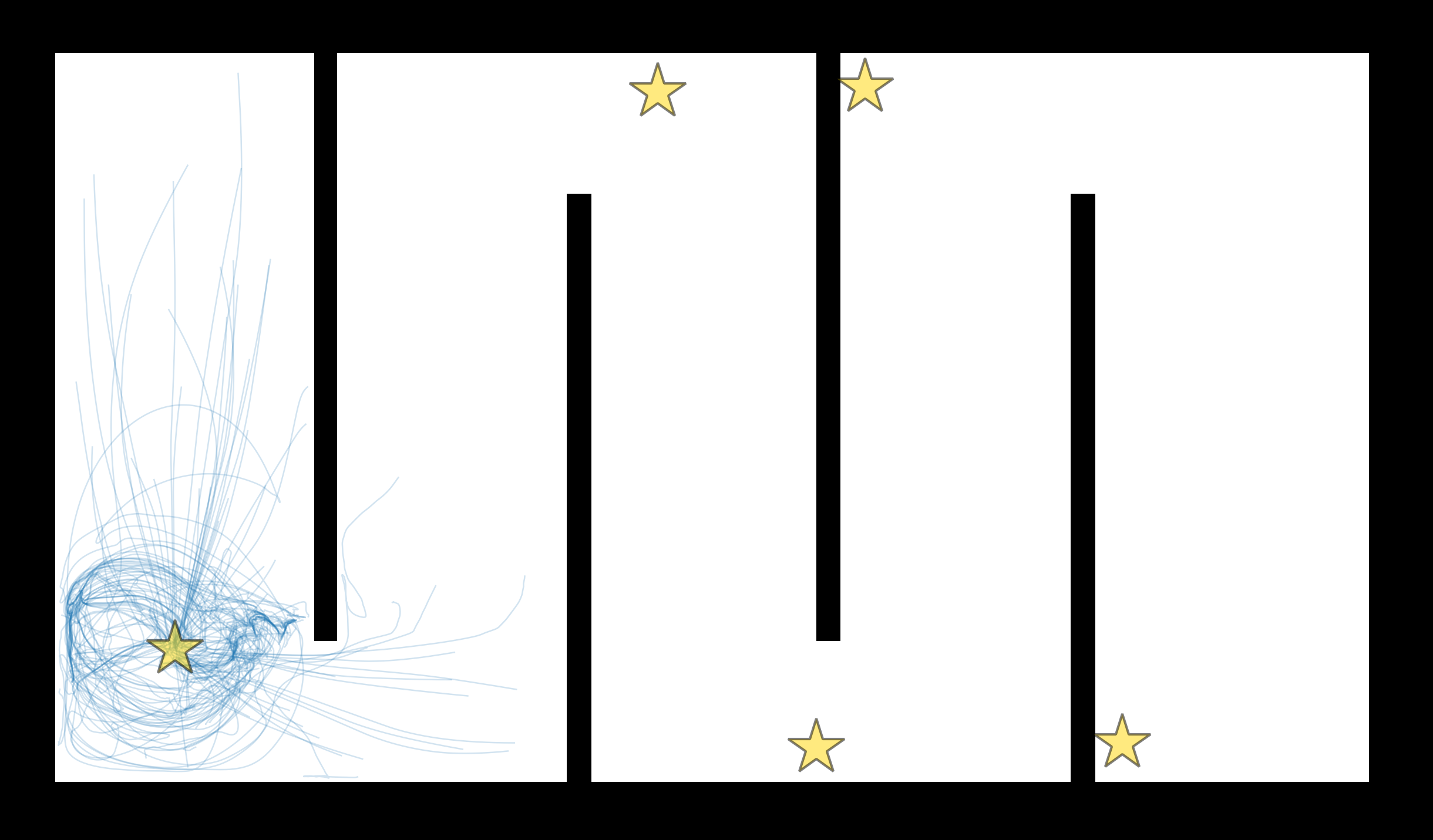}
        \label{fig:hairpin_traceplot_r3_capture}
    \end{subfigure}
    \begin{subfigure}{0.336\textwidth}
        \centering
        \vspace{-0.2cm}
        \caption{After R3 Capture $t > 11.86$s}
        \vspace{-0.2cm}
        \includegraphics[width=\linewidth]{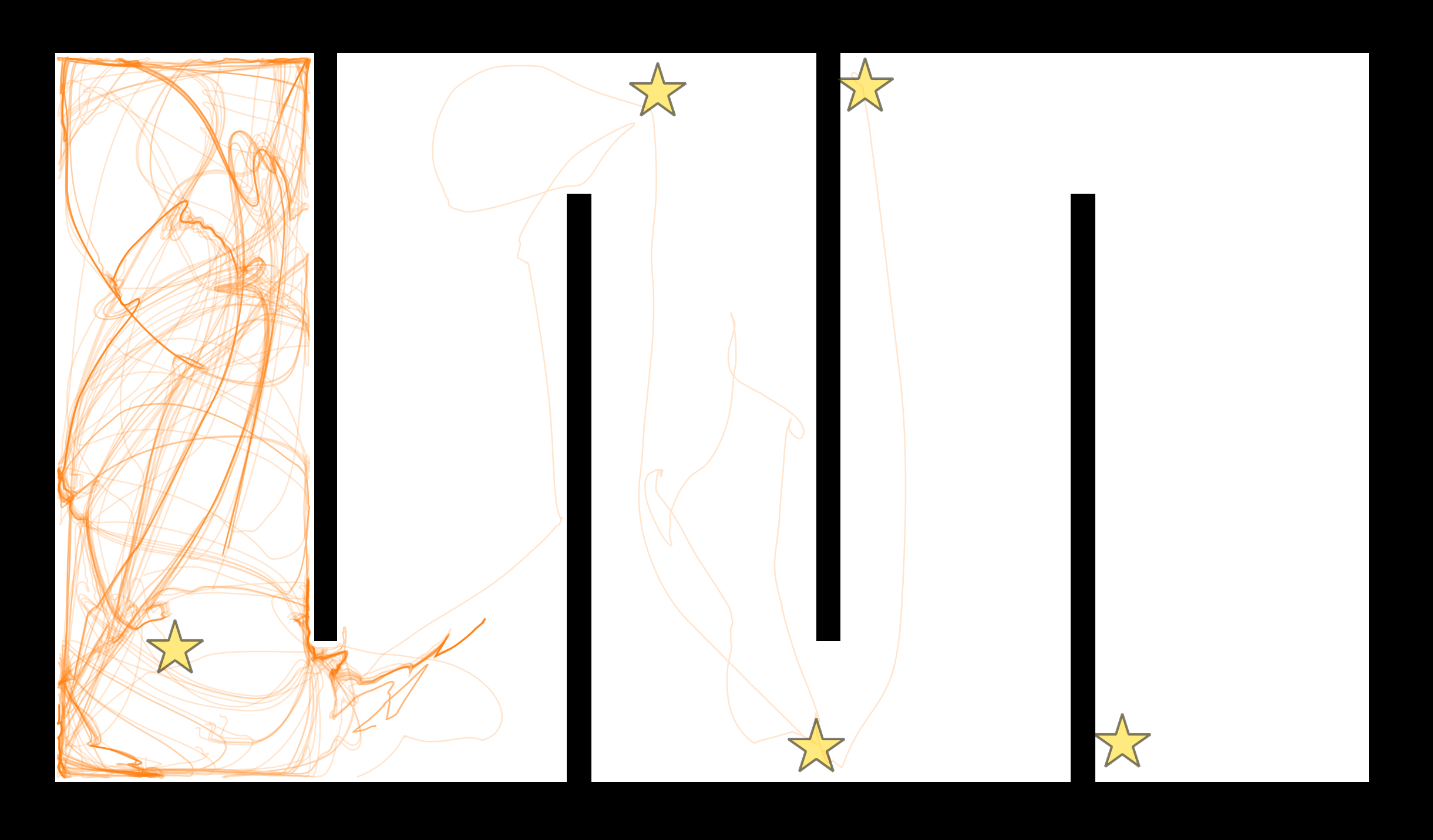}
        \label{fig:hairpin_traceplot_r3_after}
    \end{subfigure}
    \begin{subfigure}{0.336\textwidth}
        \centering
        \vspace{-0.2cm}
        \caption{R4 Capture $t \leq 18.5$s}
        \vspace{-0.2cm}
        \includegraphics[width=\linewidth]{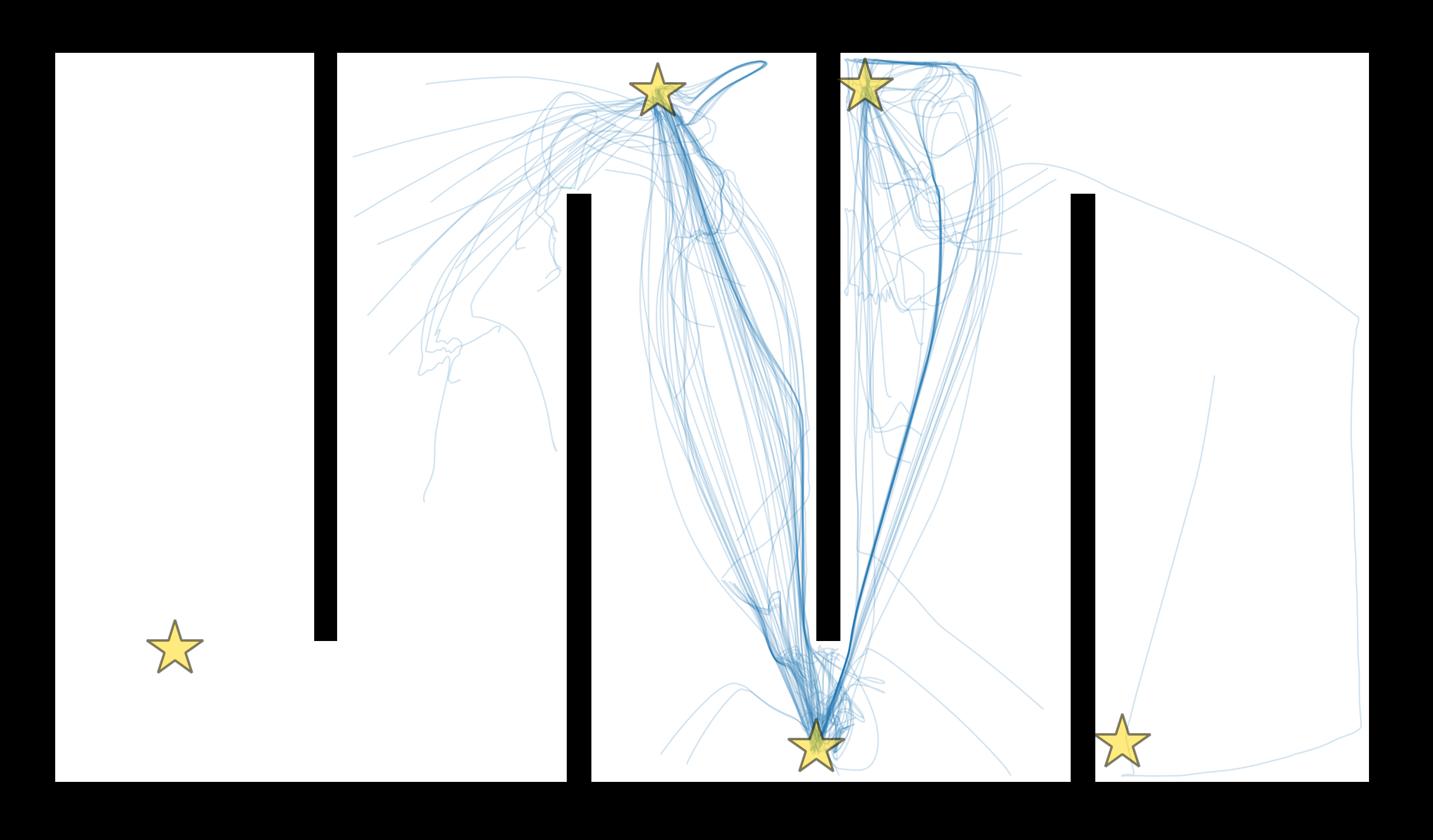}
        \label{fig:hairpin_traceplot_r4_capture}
    \end{subfigure}
    \begin{subfigure}{0.336\textwidth}
        \centering
        \vspace{-0.2cm}
        \caption{After R4 Capture $t > 18.5$s}
        \vspace{-0.2cm}
        \includegraphics[width=\linewidth]{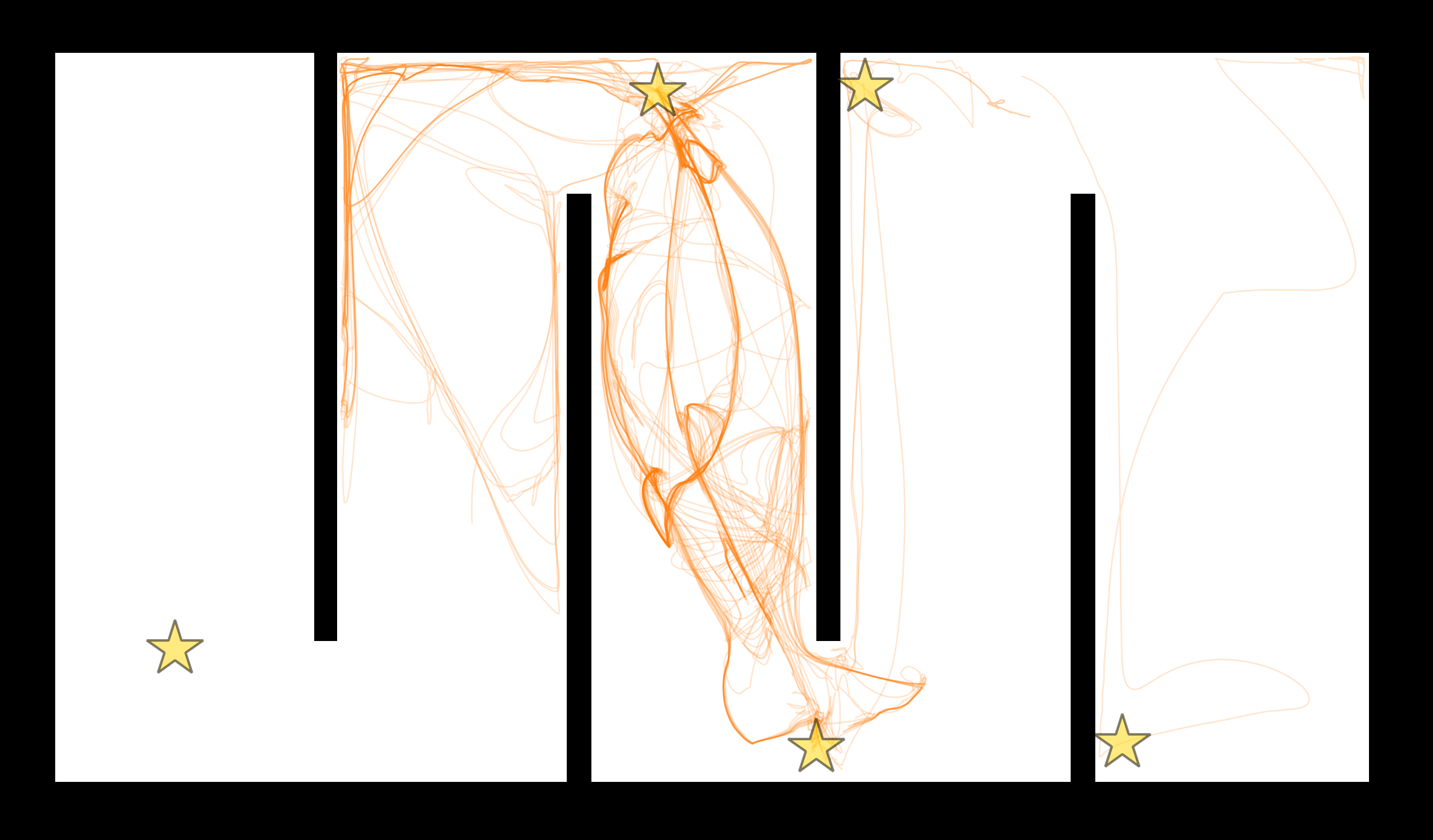}
        \label{fig:hairpin_traceplot_r4_after}
    \end{subfigure}
    \begin{subfigure}{0.336\textwidth}
        \centering
        \vspace{-0.2cm}
        \caption{R2 Capture $t \leq 25.38$s}
        \vspace{-0.2cm}
        \includegraphics[width=\linewidth]{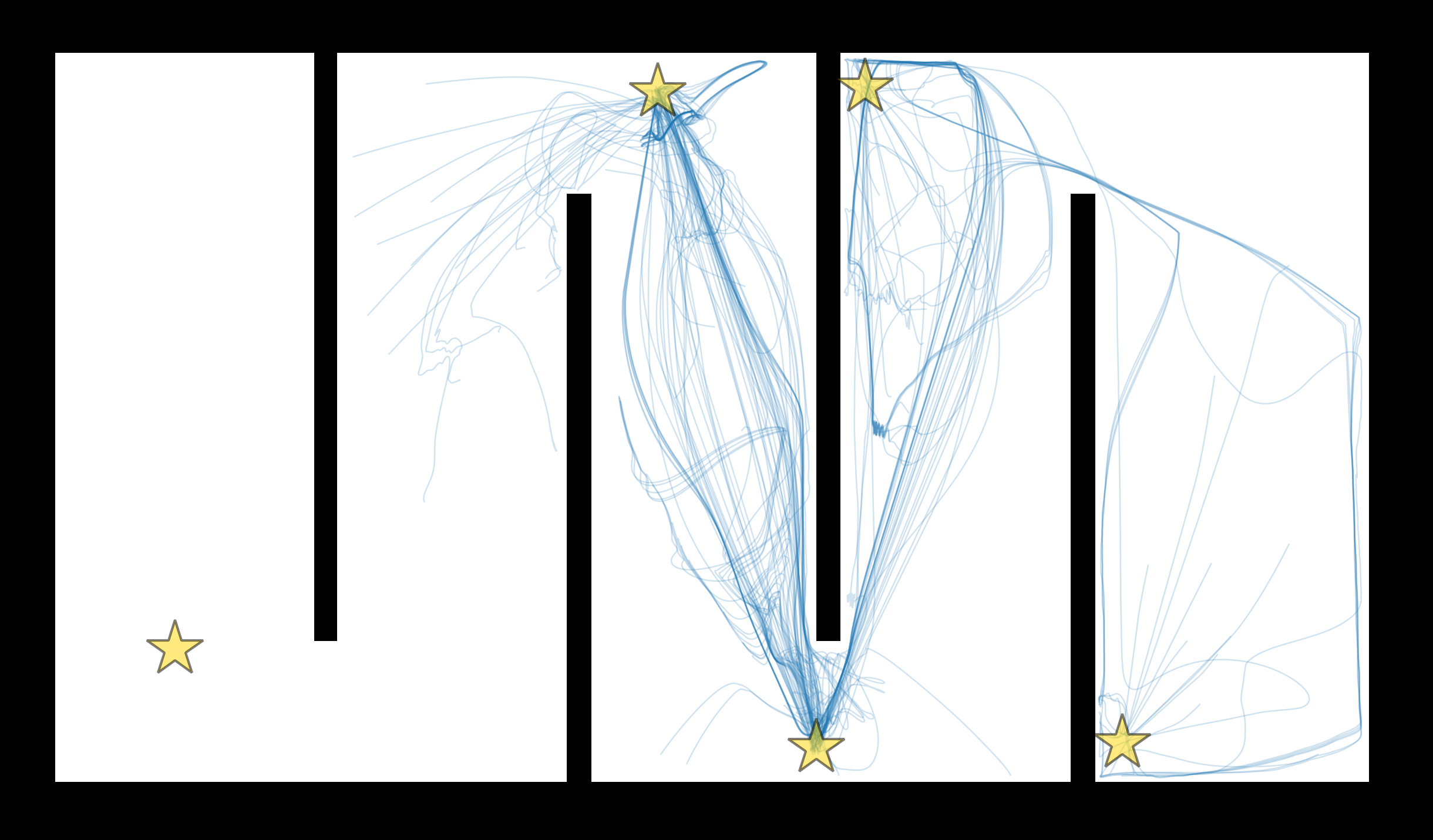}
        \label{fig:hairpin_traceplot_r2_capture}
    \end{subfigure}
    \begin{subfigure}{0.336\textwidth}
        \centering
        \vspace{-0.2cm}
        \caption{After R2 Capture $t > 25.38$s}
        \vspace{-0.2cm}
        \includegraphics[width=\linewidth]{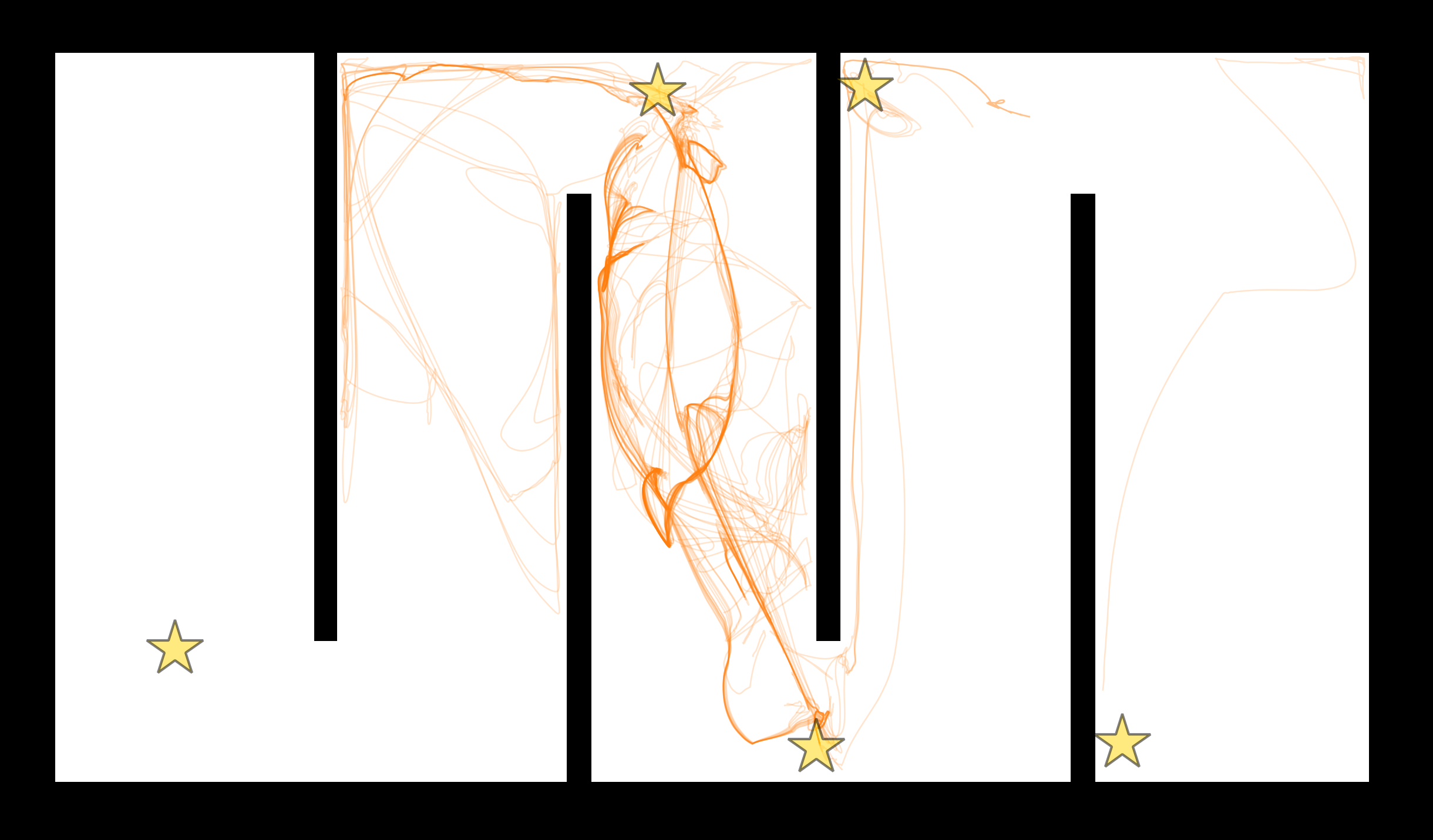}
        \label{fig:hairpin_traceplot_r2_after}
    \end{subfigure}
    \vspace{-0.5cm}
    \caption{Swarm trajectory trace in the Hairpin environment. Rewards, denoted
as R1-R5, are distributed in a counter clockwise fashion starting with the top
right-most star.}
    \label{fig:hairpin_traceplots}
\end{figure}

Next, simulations were conducted using the \ac{bo}-tuned \NS\ model
on both the hairpin and tunnel environments. Trajectory trace plots
were generated for the hairpin environment simulation, as depicted
in Figure~\ref{fig:hairpin_traceplots}. Trajectory traces in blue
depict each agent's movement throughout the simulation up to the point
of cooperative capture of a given reward, where only agents that
contributed to cooperative capture of said reward were depicted. Orange
trajectory traces reflect the behavior of the same set of agents,
after the reward had been captured. The transition from swarm goal
directed dynamics (in blue) to exploratory swarming (in orange) are
best illustrated in Figures~\ref{fig:hairpin_traceplot_r3_capture}
and~\ref{fig:hairpin_traceplot_r3_after}. A portion of the swarm collectively
collapse on the 3rd reward and then after capturing the reward immediately
disperse to look for other rewards and agents. Agents commence exploring
after capturing a reward because \NS\ does not have a global communication
system between agents, meaning that agents are not aware if other agents have
identified rewards on the other side of the map, occluded by the hairpin walls.

A key feature of our optimization process is the use of an objective function
to \textit{indirectly} evaluate swarm performance at cooperatively foraging
and capturing rewards in a distributed manner, but not to directly modify
the mechanics of the underlying swarming model. This feature allows the
objective function to evaluate swarm on collective social mechanisms, such
as distributed cooperation, but the agents of the model can be entirely
individualistic. In the case of \NS\ each agent wants to capture all 5
rewards of the hairpin environment, even if other agents have already
captured said reward. Figures~\ref{fig:hairpin_traceplot_r2_capture}
and~\ref{fig:hairpin_traceplot_r2_after} illustrate this scenario, where reward
2 is the last reward to be captured (top center-left), meaning that collectively
the swarm has cooperatively captured all 5 rewards. However, since the agents
are individually trying to capture all 5 rewards and they have not done so at $t
= $25.38s, some agents are shown in Figure~\ref{fig:hairpin_traceplot_r2_after}
going back to capture rewards 5 (bottom center) and 1 (top center right).

\begin{figure}[tb!]
    \centering
    \begin{subfigure}{0.36\textwidth}
        \centering
        \vspace{-0.2cm}
        \caption{R2 Capture $t \leq 0.95$s}
        \vspace{-0.2cm}
        \includegraphics[width=\linewidth]{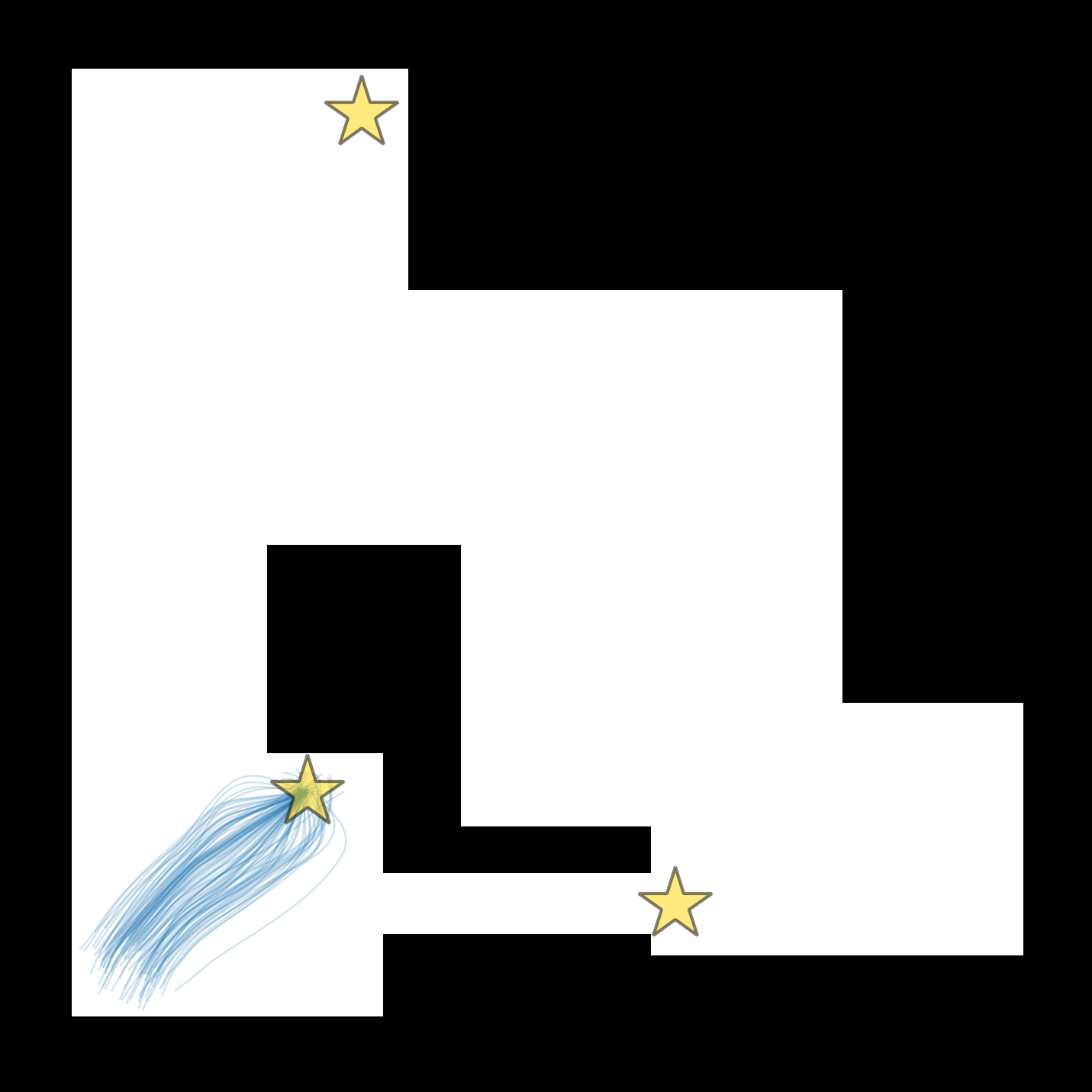}
        \label{fig:tunnel_traceplot_r2_capture}
    \end{subfigure}
    \begin{subfigure}{0.36\textwidth}
        \centering
        \vspace{-0.2cm}
        \caption{After R2 Capture $t > 0.95$s}
        \vspace{-0.2cm}
        \includegraphics[width=\linewidth]{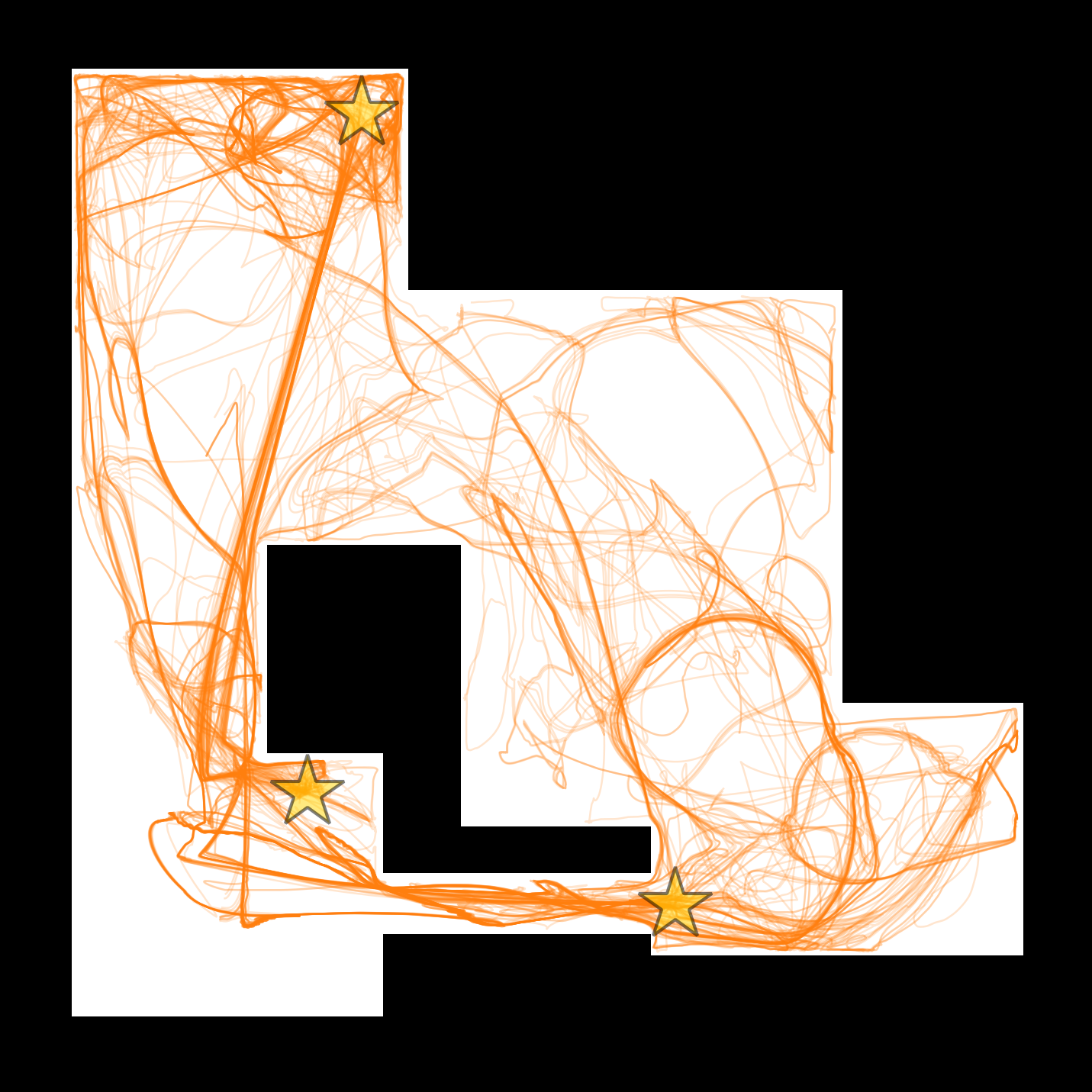}
        \label{fig:tunnel_traceplot_r2_after}
    \end{subfigure}
    \begin{subfigure}{0.36\textwidth}
        \centering
        \vspace{-0.2cm}
        \caption{R1 Capture $t \leq 5.46$s}
        \vspace{-0.2cm}
        \includegraphics[width=\linewidth]{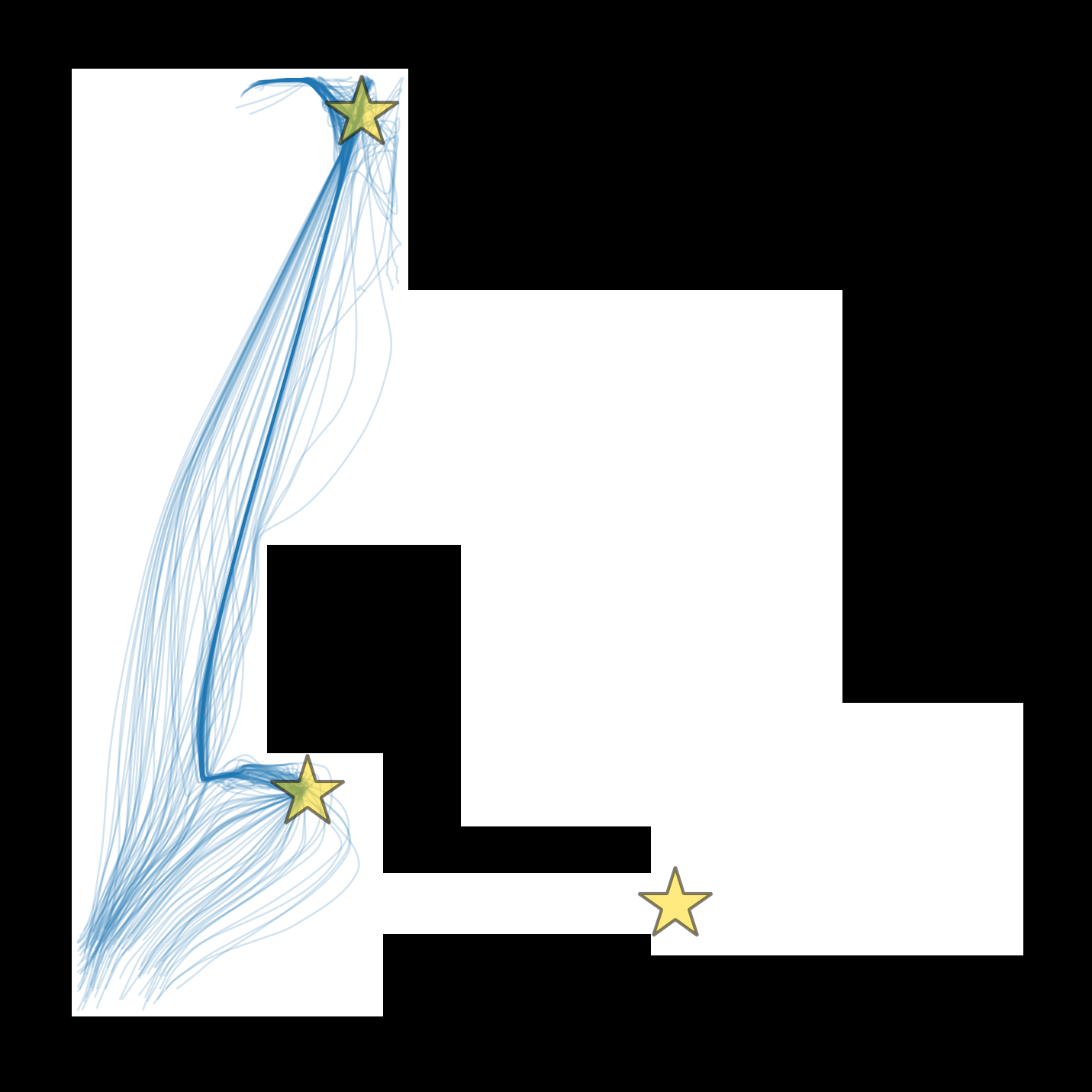}
        \label{fig:tunnel_traceplot_r1_capture}
    \end{subfigure}
    \begin{subfigure}{0.36\textwidth}
        \centering
        \vspace{-0.2cm}
        \caption{After R1 Capture $t > 5.46$s}
        \vspace{-0.2cm}
        \includegraphics[width=\linewidth]{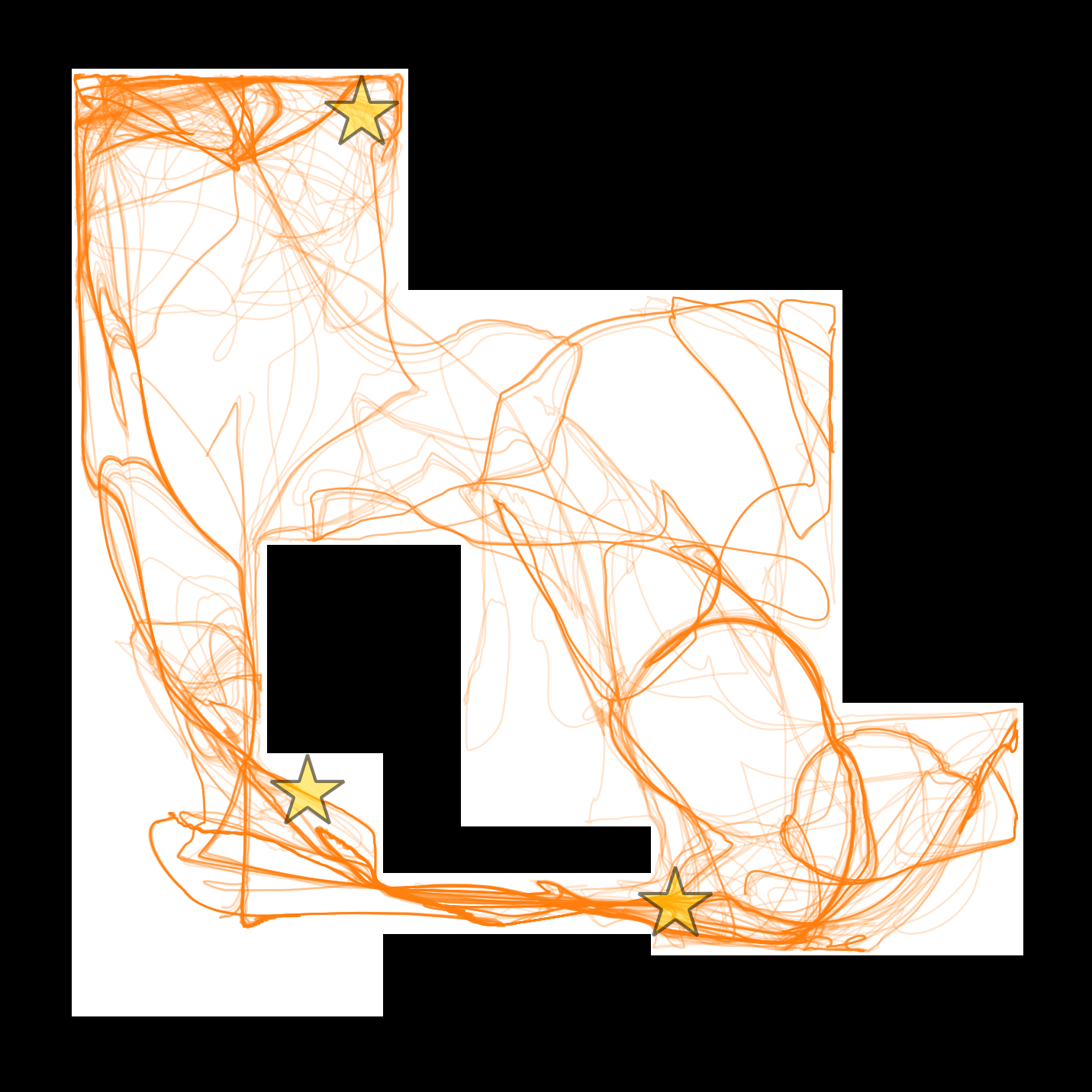}
        \label{fig:tunnel_traceplot_r1_after}
    \end{subfigure}
    \begin{subfigure}{0.36\textwidth}
        \centering
        \vspace{-0.2cm}
        \caption{R3 Capture $t \leq 31.78$s}
        \vspace{-0.2cm}
        \includegraphics[width=\linewidth]{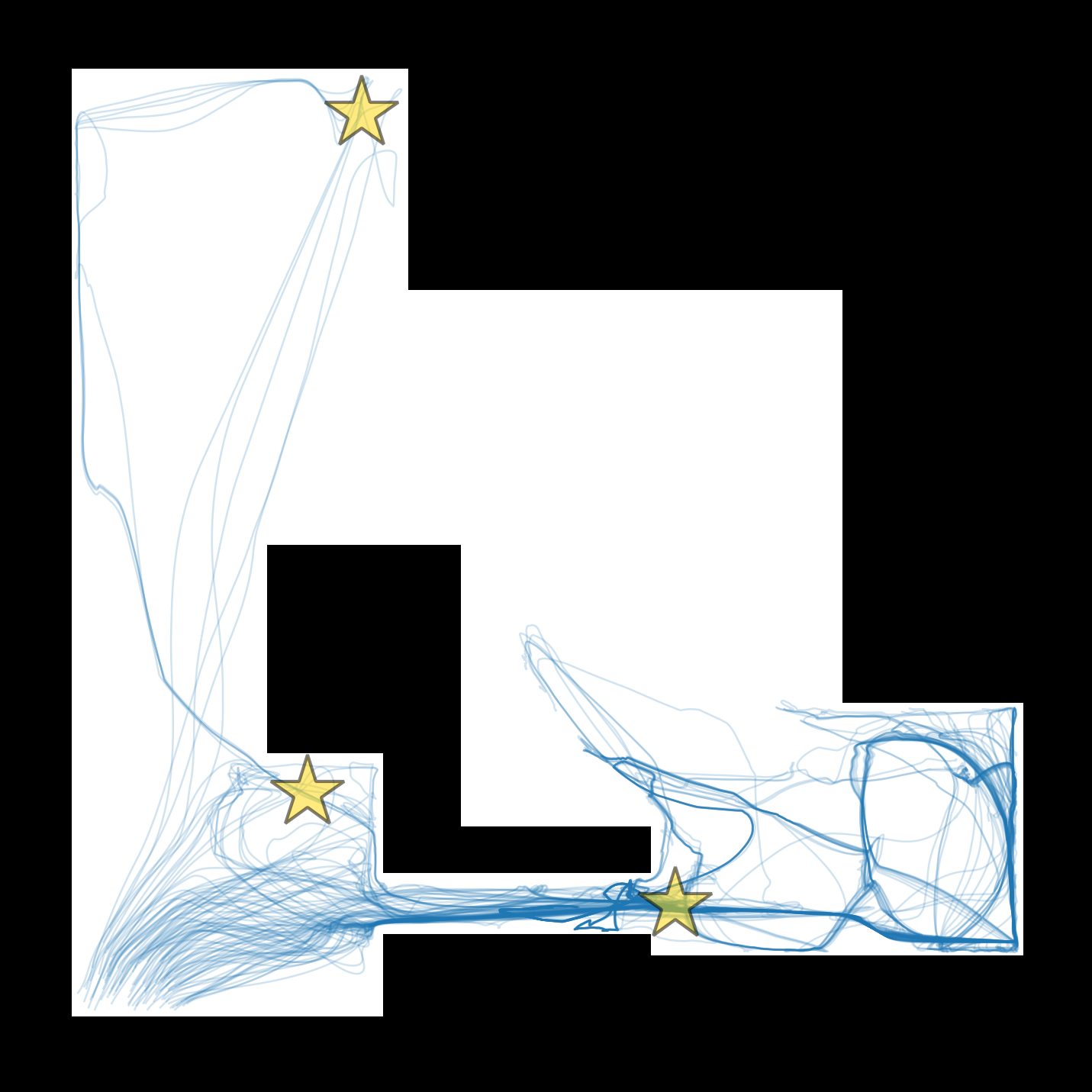}
        \label{fig:tunnel_traceplot_r3_capture}
    \end{subfigure}
    \begin{subfigure}{0.36\textwidth}
        \centering
        \vspace{-0.2cm}
        \caption{After R3 Capture $t > 31.78$s}
        \vspace{-0.2cm}
        \includegraphics[width=\linewidth]{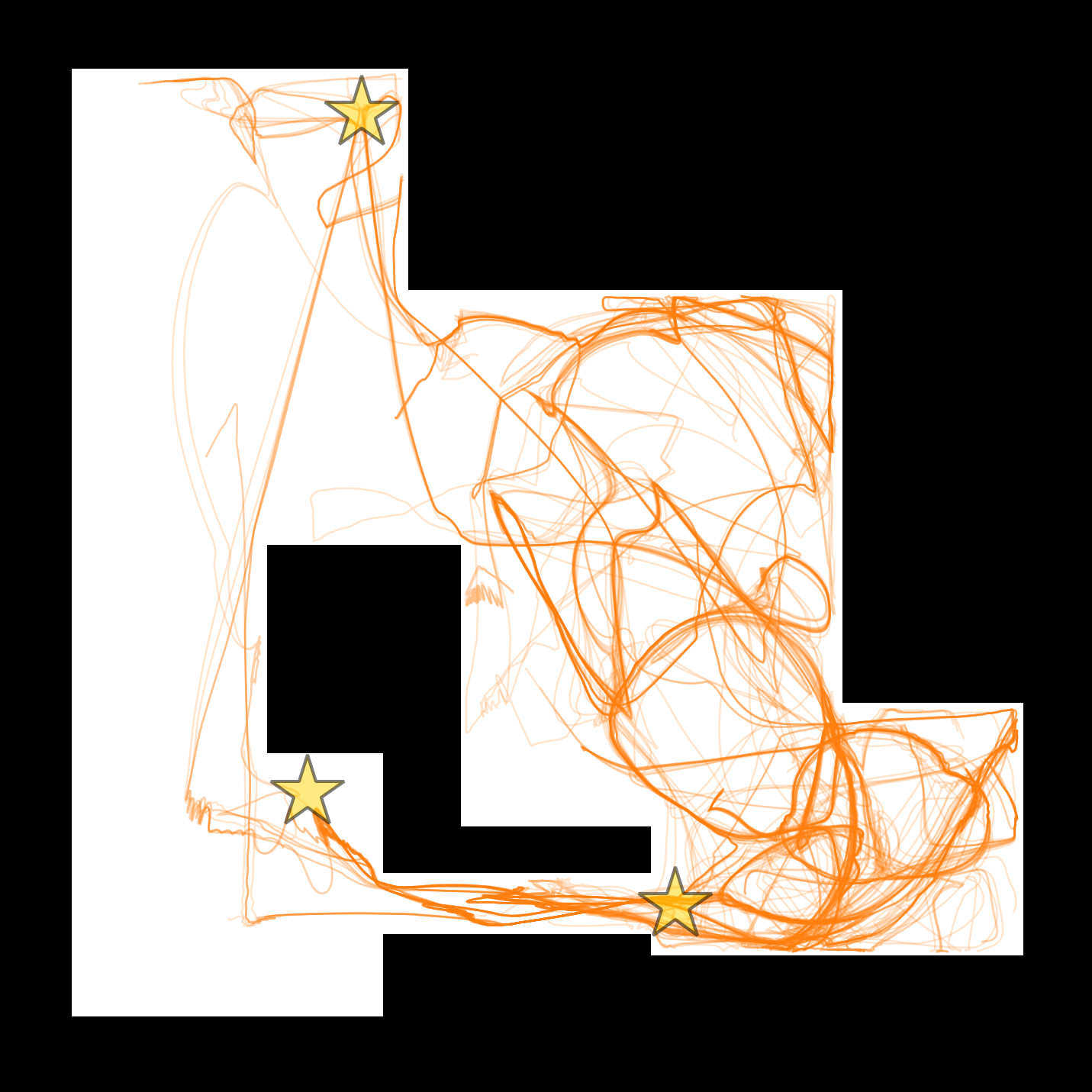}
        \label{fig:tunnel_traceplot_r3_after}
    \end{subfigure}
    \vspace{-0.5cm}
    \caption{Swarm trajectory trace in the Tunnel environment. Rewards, denoted
as R1-R3, are distributed in a counter clockwise fashion starting with the top
left-most star.}
    \label{fig:tunnel_traceplots}
\end{figure}

The tunnel environment, shown in Figure~\ref{fig:tunnel_traceplots}, provides
a scenario whereby the swarm must distribute in order to forage. Unlike
the hairpin environment which uniformly at random spawns agents across
the entire environment, the tunnel environment spawns all agents in the
bottom left corner. As a result, the entire swarm collectively captured a
single reward (reward 2 in Figure~\ref{fig:tunnel_traceplot_r2_capture}
and then had to immediately bifurcate in order to successfully capture the
remaining two rewards, see Figures~\ref{fig:tunnel_traceplot_r1_capture}
and~\ref{fig:tunnel_traceplot_r3_capture}. This scenario illustrates the success
of the parameter optimization process in selecting parameters that encourage
flexible swarming that allows a swarm to work in unison and then dissolve in
order to seek further rewards. A challenging feature of the tunnel environment
is that reward 3 (bottom right) is visible from the spawn position (bottom left)
and closer than reward 2, yet has limited accessibility due to the constricting
tunnel. Reward 2, on the other hand, is easily accessible yet further away and
somewhat occluded after agents transition to reward 1's (bottom left) position.
The substantially faster capture time of reward 1 (5.46s) vs reward 2's much
slower 31.78s illustrates that the parameter choice indirectly prioritized
ease of exploration over efficiency in reward capture. Moreover, comparing the
before (blue) and after (orange) capture of each of the rewards we identified
that agents began to use the large opening in the center of the map only once
enough agents had reached the bottom right corner of the map. This phenomenon
characterizes the parameter choice as encouraging exploration only once the
sub-swarm reaches sufficient membership.

\subsection{Exploring Future Parameter Space}


\begin{figure*}
    \centering
    \includegraphics[width=\linewidth,trim=12cm 8cm 14cm 3cm,clip]{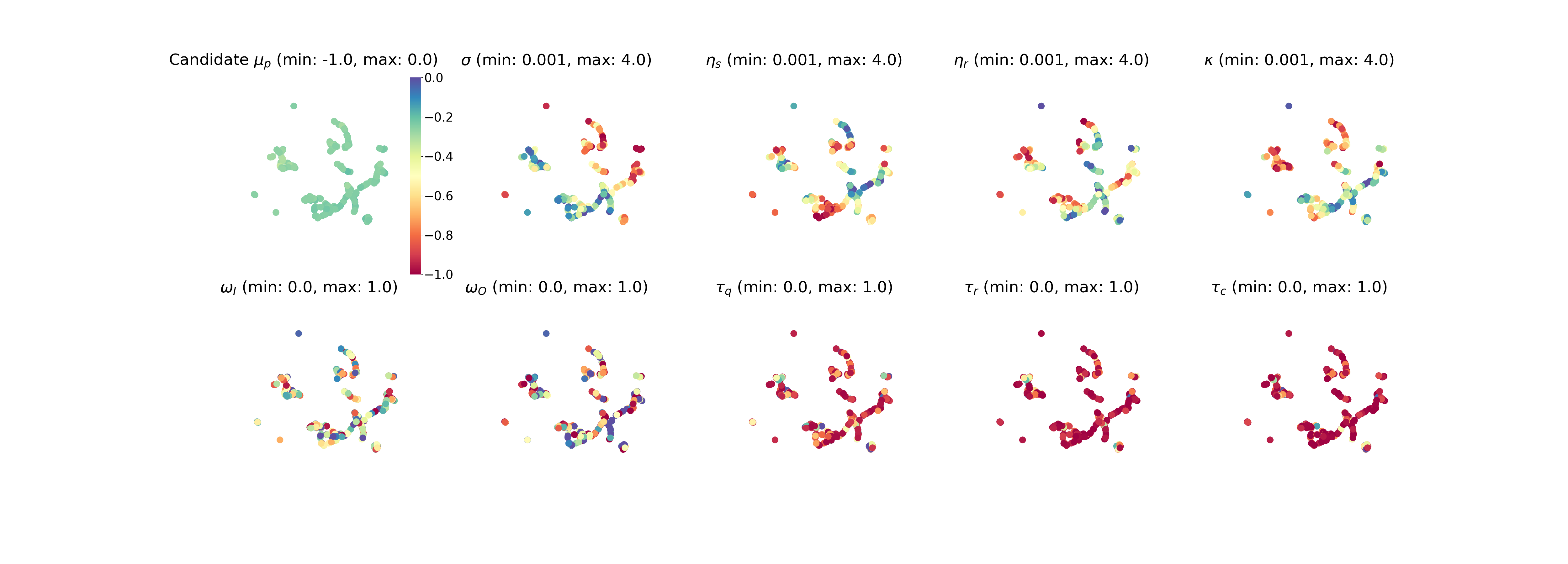}
    \caption{\ac{ei} anticipated parameter space}
    \label{fig:acq_sample_qEI_params}
\end{figure*}


An additional benefit to intelligently exploring the parameter space with acquisition functions is that these trained acquisition functions can then be used to predict the performance of unsimulated regions of the parameter space.
In Figure~\ref{fig:acq_sample_qEI_params} we depict 500 samples from the parameter space chosen by the \ac{ei} acquisition function.
The samples were evaluated using the trained \ac{gp} model's posterior distribution corresponding to the \ac{ei} acquisition function.
The posterior means (shown in the top left plot) are similar for most points since the \ac{ei} acquisition function is targeting regions of the parameter space that are the most likely to produce improvement, which results in samples that are perceived to produce a high objective value (near 0).
Following a similar procedure to Subsection~\ref{ssec:qual_eval}, we selected a set of candidate points from this \ac{ei} explored parameter space and simulated the set of parameters on the hairpin and tunnel environments.

\begin{figure}[tb!]
    \centering
    \begin{subfigure}{0.36\textwidth}
        \centering
        \vspace{-0.2cm}
        \caption{Hairpin Reward Capture}
        \vspace{-0.2cm}
        \includegraphics[width=\linewidth]{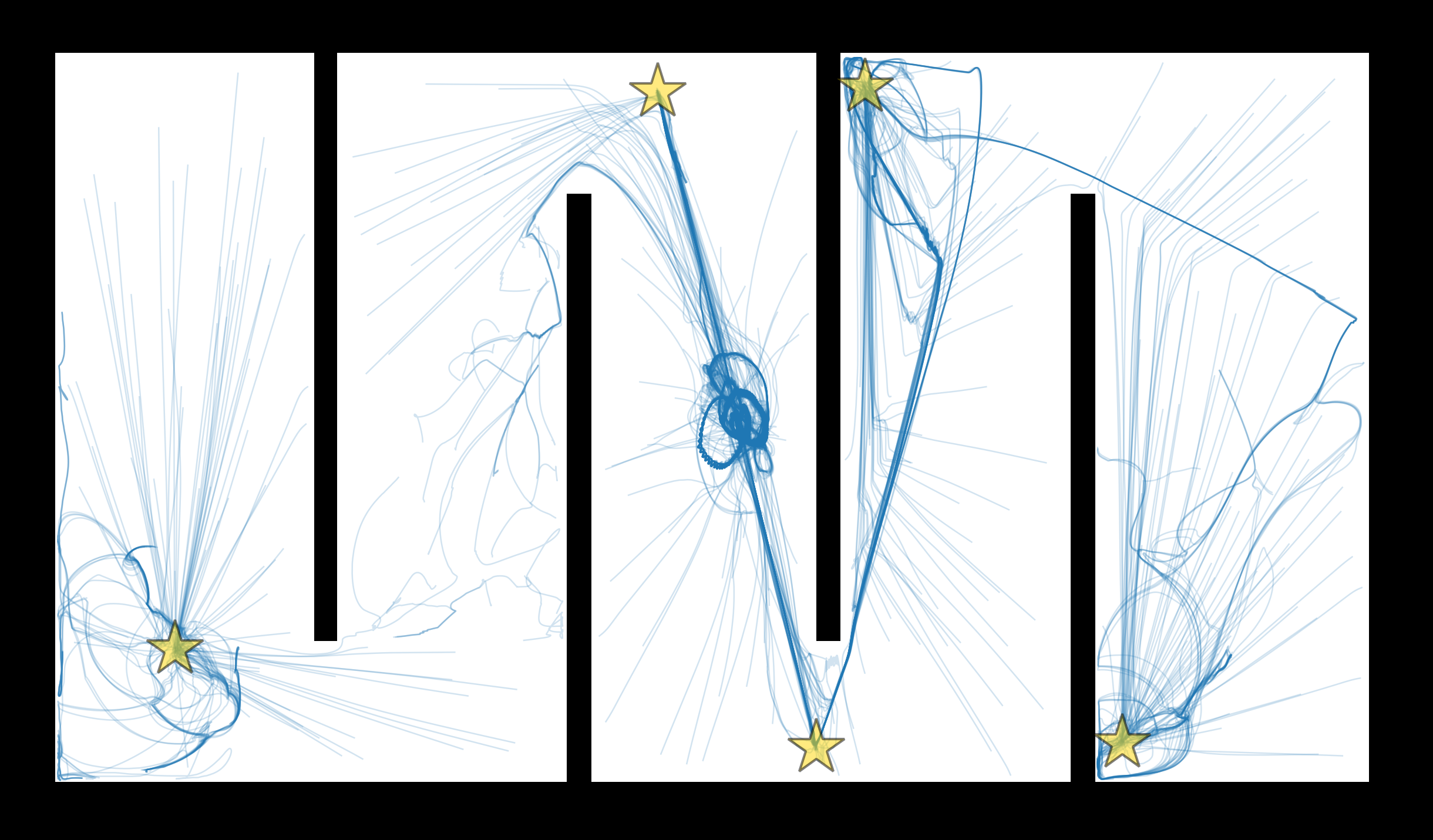}
        \label{fig:acq_qei_hairpin}
    \end{subfigure}
    \begin{subfigure}{0.36\textwidth}
        \centering
        \vspace{-0.2cm}
        \caption{Hairpin After Capture $t > 47.44$s}
        \vspace{-0.2cm}
        \includegraphics[width=\linewidth]{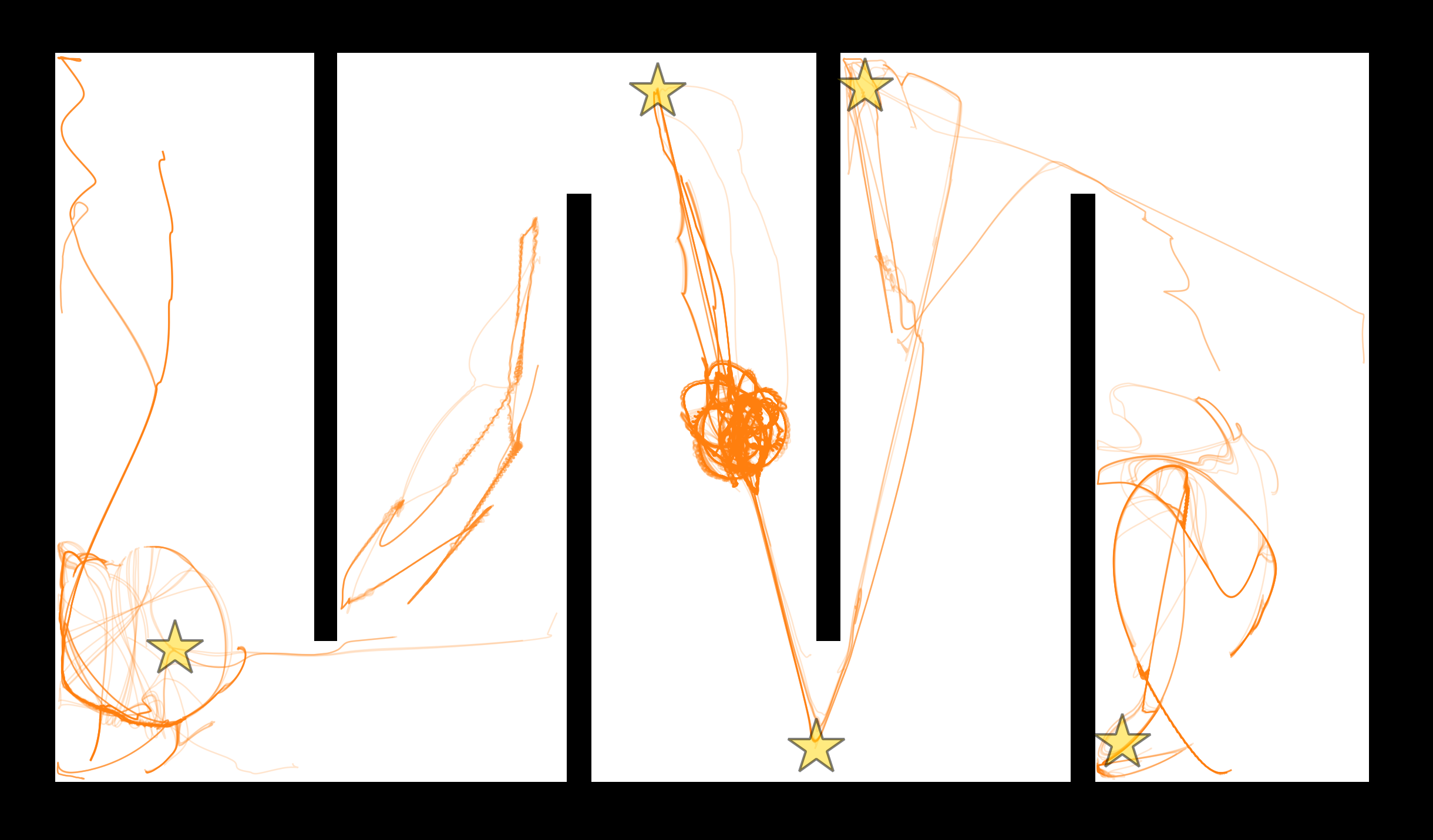}
        \label{fig:acq_qei_hairpin2}
    \end{subfigure}
    \begin{subfigure}{0.36\textwidth}
        \centering
        \vspace{-0.2cm}
        \caption{Tunnel Reward Capture}
        \vspace{-0.2cm}
        \includegraphics[width=\linewidth]{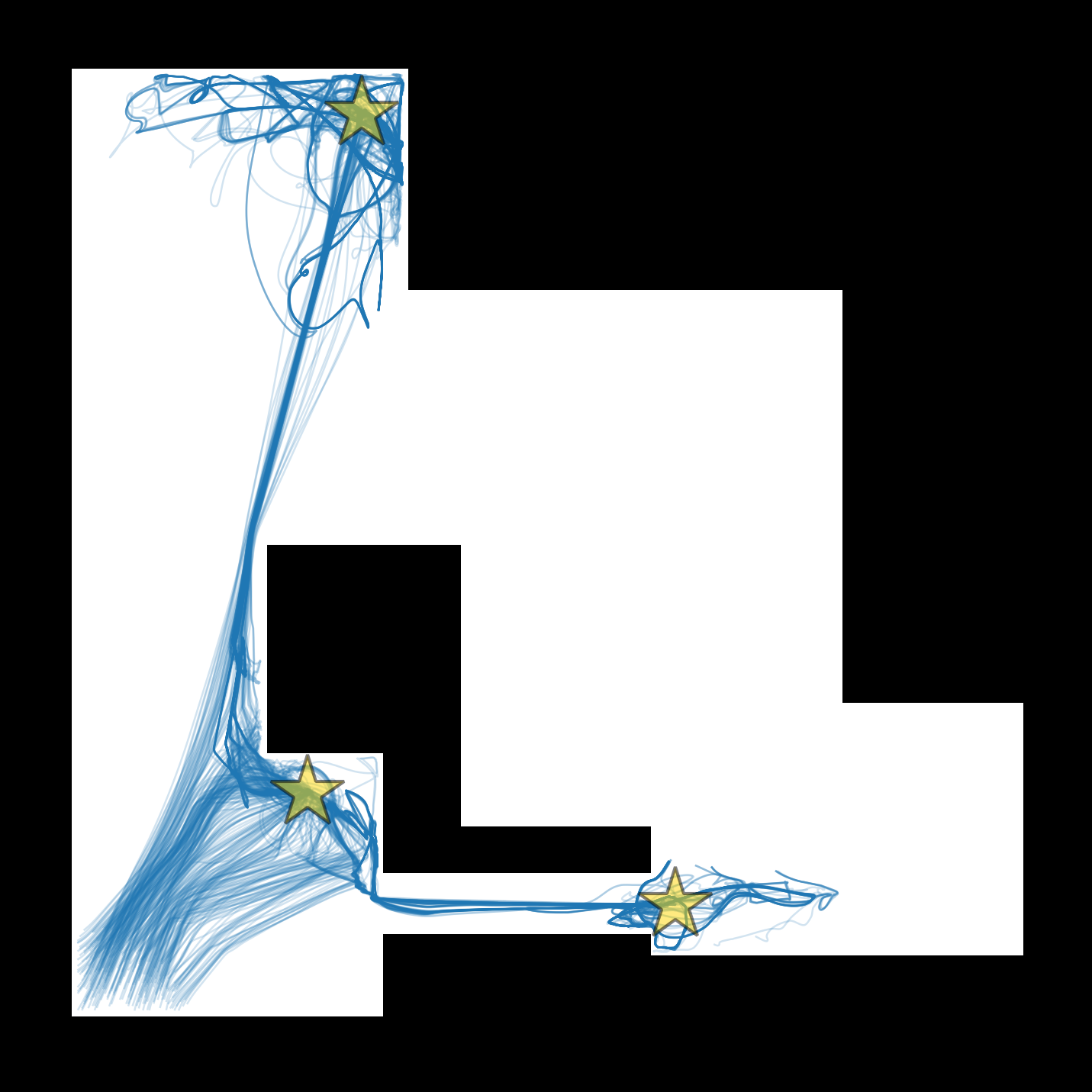}
        \label{fig:acq_qei_tunnel}
    \end{subfigure}
    \begin{subfigure}{0.36\textwidth}
        \centering
        \vspace{-0.2cm}
        \caption{Tunnel After Capture $t > 66.96$s}
        \vspace{-0.2cm}
        \includegraphics[width=\linewidth]{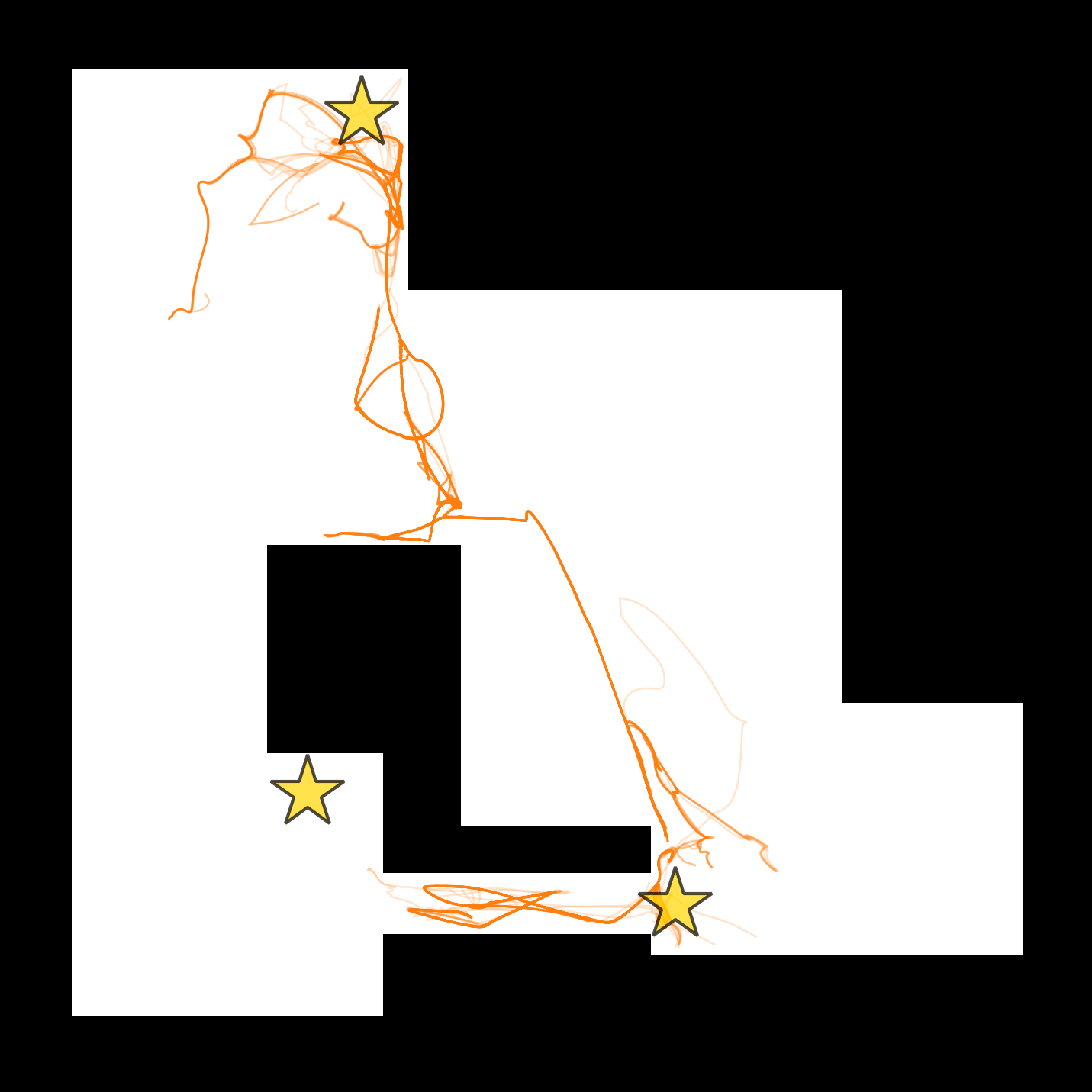}
        \label{fig:acq_qei_tunnel2}
    \end{subfigure}
    \vspace{-0.5cm}
    \caption{Swarm trajectory trace in the hairpin and tunnel environments using
parameters selected from \ac{ei} anticipated parameter space.}
    \label{fig:acq_traceplots}
\end{figure}

We selected a set of parameters that were predicted to be high performing
according to the \ac{gp} model's posterior mean and that contained unique
selections rarely seen in the parameter space. In particular, we were able to
find a combination of parameters where $\tau_q, \tau_r > 1^{-3}$, which is rare
since small time constants allow for faster responses which are optimal for
fast reward capture. The $\sigma$ and $\kappa$ the parameters corresponding
to exploration were also not at extremes but instead were mid range values.
Last, $\eta_s$ and $\eta_r$ were not in direct contrast with one another,
but were both mid-range values which may be sub-optimal since agents will
learn with respect to inter-agent dynamics and reward response will be less
impactful. Figure~\ref{fig:acq_traceplots} shows trajectory traces of all
agents before and after all rewards were cooperatively captured by the swarm
on the hairpin and tunnel environments. While all rewards were successfully
cooperatively captured by the swarm, as the \ac{gp} model posterior predicted,
the selected set of parameters resulted in slower reward capture. The traceplots
show slower (47.44s for hairpin and 66.96s for tunnel) reward capture on
both environments compared with the more optimal parameter choice shown in
Figures~\ref{fig:hairpin_traceplots} and~\ref{fig:tunnel_traceplots} (25.38s for
hairpin 31.78s for tunnel). The slower reward capture is likely attributed to
the larger $\tau_q = 0.85$ and $\tau_r = 0.55$, which result in slower recurrent
and reward updates. However, the swarm exhibited more deliberate cooperation
with substantially less erratic exploratory behavior which coincides with
neither $\kappa$ or $\sigma$ being set to the upper limit. In the case where
energy consumption related to movement was taken into account, this slower set
of parameters would be advantageous as it exhibited effective cooperation with
minimal unnecessary exploration.

\section{Conclusion}
\label{sec:conclusion}

Neuroscience-inspired learning and control methods have become of increasing
interest in the domains of Robotics and Artificial Intelligence, in particular
for multi-agent control. Here, we presented a method of exploring and
visualizing the parameter space of a multi-agent model in a sample efficient
manner using \ac{bo}. We introduced an objective function for quantifying
emergent cooperative foraging for a multi-agent model with no inherent
goal-directed cooperative mechanics. Using this objective function, our trained
\ac{gp} model was able to correctly predict the performance of the complex \NS\
model on the task of cooperative foraging across two distinct environments and
without simulation on $N_p$ (9) tunable parameters. Training the surrogate
\ac{gp} model was facilitated by directed exploration of the parameter space of
the underlying complex model through the use of the qEI or qNoisyEI acquisition
functions. The qEI acquisition function was shown to direct exploration towards
parameter sets that maximized utility, even over hand-tuned default parameters.
Through the use of \ac{umap}~\cite{mcinnes2018umap} we demonstrated a means to
visualizing an $N_p$-dimensional parameter space to identify and select high
performing sets of parameters. Lastly, we illustrated the capability of our
approach to successfully identify sets of parameters that generalize across
environments through evaluation on two environments using the same set of
parameters. Overall, our approach serves as a framework by which parameters of
complex multi-agent models can be explored and selected to maximize cooperative
foraging and reward capture.

In future work, we will consider exploring objective functions that are agnostic
to the characteristics of environments (e.g. rewards). Instead, we postulate
that the variation of swarm spatial structure can serve as means to evaluate the
level of cooperation of a swarm in meeting the goals of an environment, without
knowledge of the goals themselves. Such an objective function could extend the
flexibility of our approach to operate in any environment and to tasks that
exhibit difficult to quantify goals.

\section{Acknowledgments}
\label{sec:acknowledgments}

This work was supported by NSF award NCS/FO 1835279 to GMH, KZ, KMS, and JDM;
JHU/APL internal research and development awards to AH, GMH, and KMS; and
the JHU/Kavli Neuroscience Discovery Institute and JHU/APL Innovation and
Collaboration Janney Program provided additional support to GMH.


\bibliographystyle{elsarticle-num}
\bibliography{cas-refs}

\newpage
\begin{appendices}
\label{sec:appendix}

In this section we provide additional technical details on parameters not
investigated and observations on the default set of parameters.

\section{Constants}
Model parameters held constant and used by all \NS\ models are described in
Table~\ref{tbl:constants}. Additionally, the \ac{bo} process across all three
acquisition functions included the following constants: 256 \ac{mc} samples,
30 training epochs with a batch size of 3, and 8 initial training examples to
initialize the \ac{gp} model.

\begin{table}[htb!]
\scriptsize
    \begin{center}
        \ra{1.3}
        \begin{tabular}{@{} lrl @{}}
        \toprule
            Parameter & Range & Description \\
        \midrule
            $N_s$ & $300$ & Number of agents \\
            $\Delta t$ & $0.01$ & Simulation integration time step in $s$ \\
            Duration & $200$ & Simulation duration in $s$ \\
            $E_{\text{max}}$ & $3\text{e}3$ & Maximum kinetic energy in $\frac{kg \text{ points}^2}{s^2}$ \\
            $\mu_{m}$ & $0.9$ & Momentum coefficient in $kg$ \\
            $g_s$ & $0.5$ & Swarming inputs gain \\
            $g_r$ & $0.3$ & Reward inputs gain \\
            $g_c$ & $0.2$ & Sensory cue inputs gain \\
            $d_{text{rad}}$ & $12$ & Reward contact radius for capture \\
        \bottomrule
        \end{tabular}
    \end{center}
\caption{\NS\ model constants.}
\label{tbl:constants}
\end{table}

\section{Default Parameters}
\begin{figure}[tb!]
    \centering
    \begin{subfigure}{0.336\textwidth}
        \centering
        \caption{R3 Capture $t \leq 3.19$s}
        \vspace{-0.2cm}
        \includegraphics[width=\linewidth]{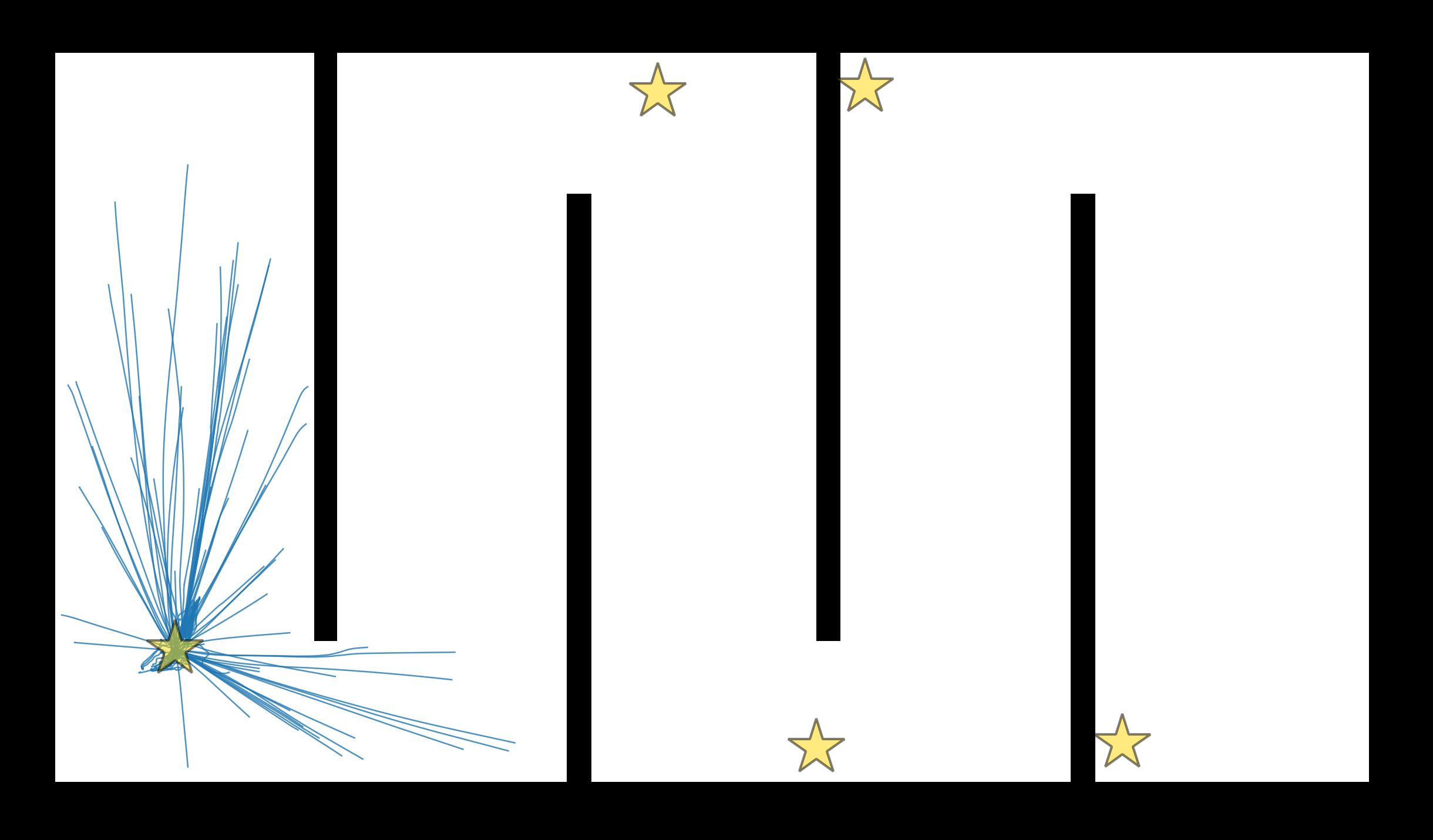}
        \label{fig:default_hairpin_traceplot_r3_capture}
    \end{subfigure}
    \begin{subfigure}{0.336\textwidth}
        \centering
        \caption{After R3 Capture $t > 3.19$s}
        \vspace{-0.2cm}
        \includegraphics[width=\linewidth]{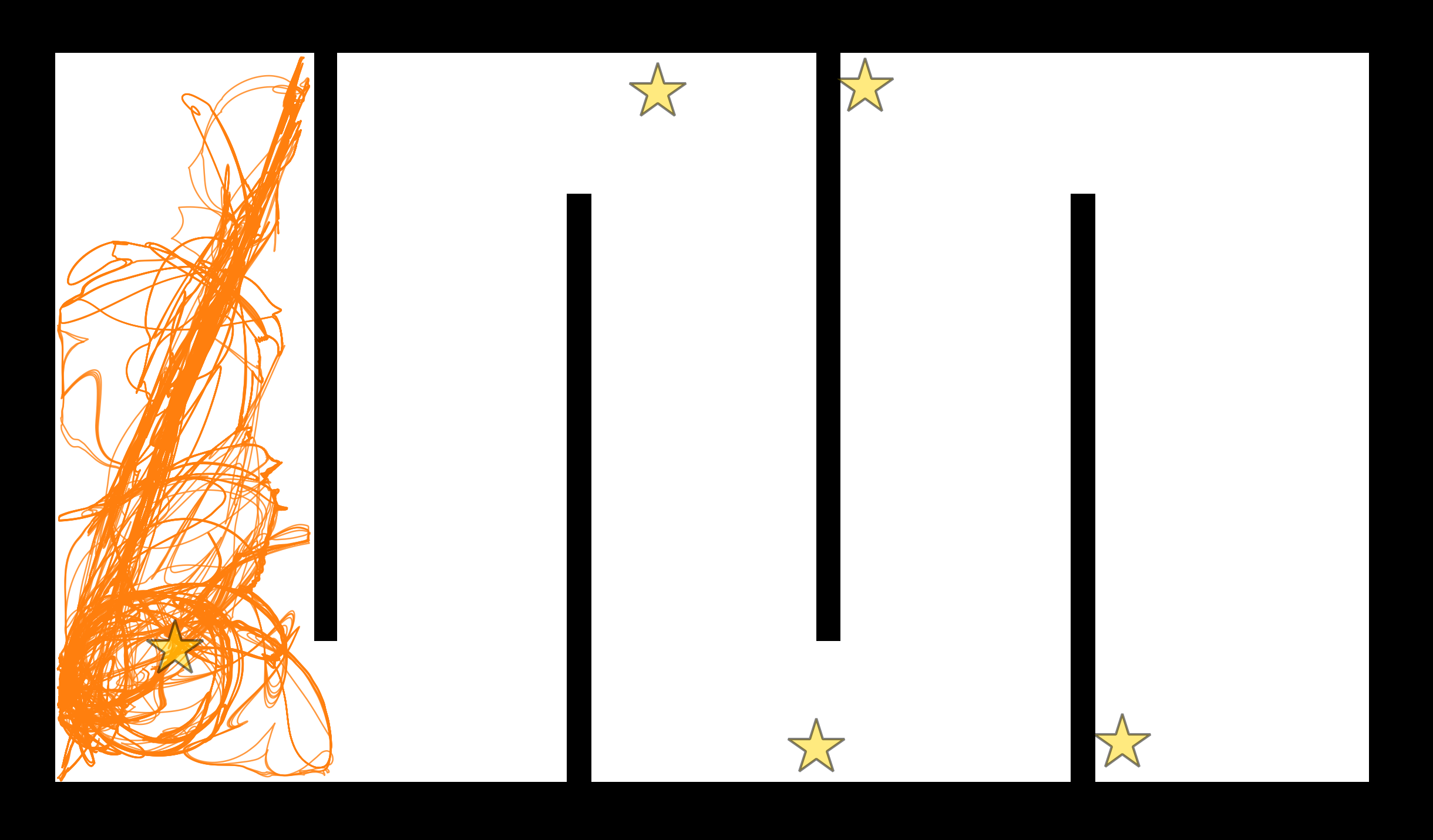}
        \label{fig:default_hairpin_traceplot_r3_after}
    \end{subfigure}
    \begin{subfigure}{0.336\textwidth}
        \centering
        \vspace{-0.2cm}
        \caption{R5 Capture $t \leq 5.15$s}
        \vspace{-0.2cm}
        \includegraphics[width=\linewidth]{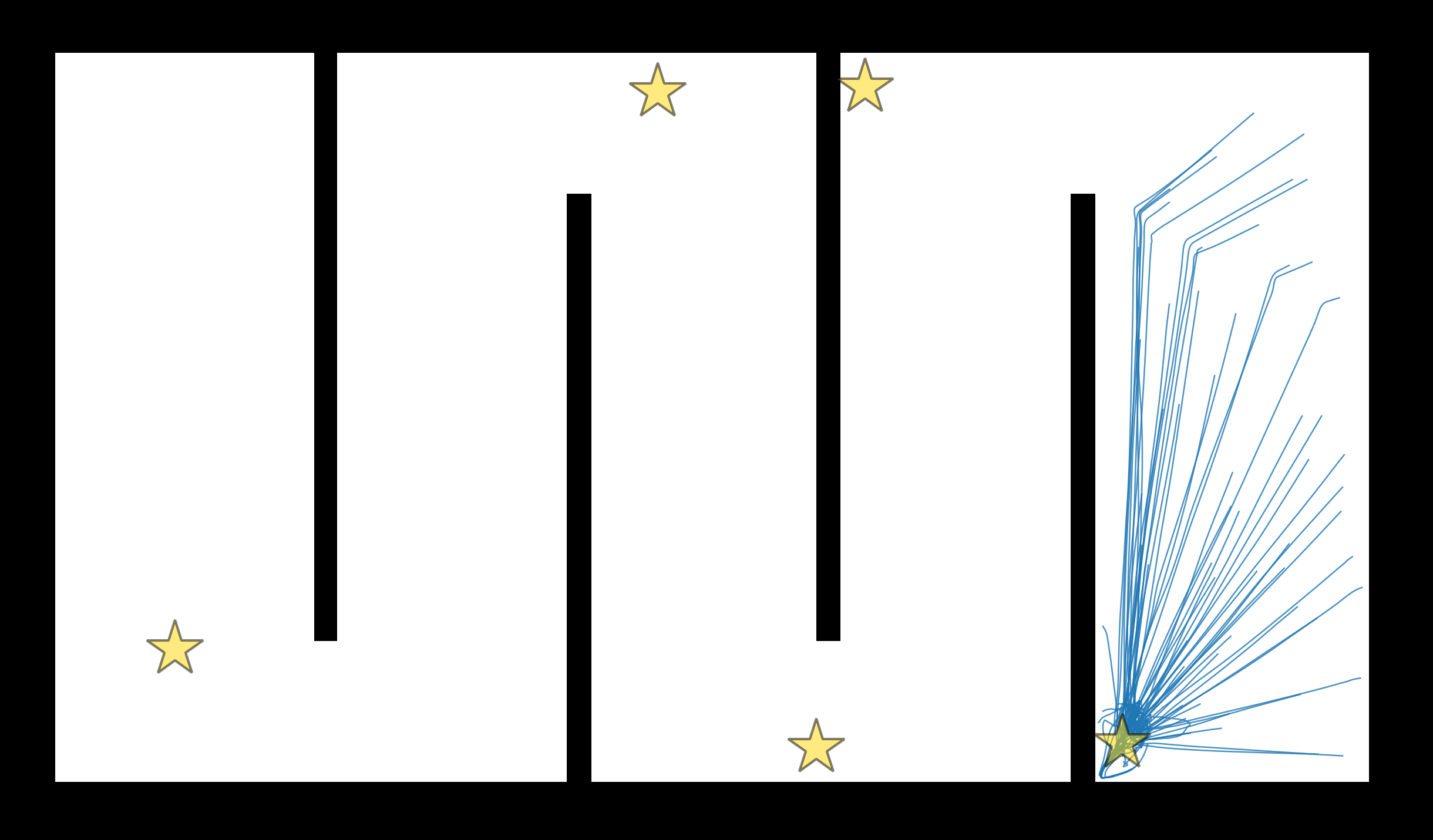}
        \label{fig:default_hairpin_traceplot_r5_capture}
    \end{subfigure}
    \begin{subfigure}{0.336\textwidth}
        \centering
        \vspace{-0.2cm}
        \caption{After R5 Capture $t > 5.15$s}
        \vspace{-0.2cm}
        \includegraphics[width=\linewidth]{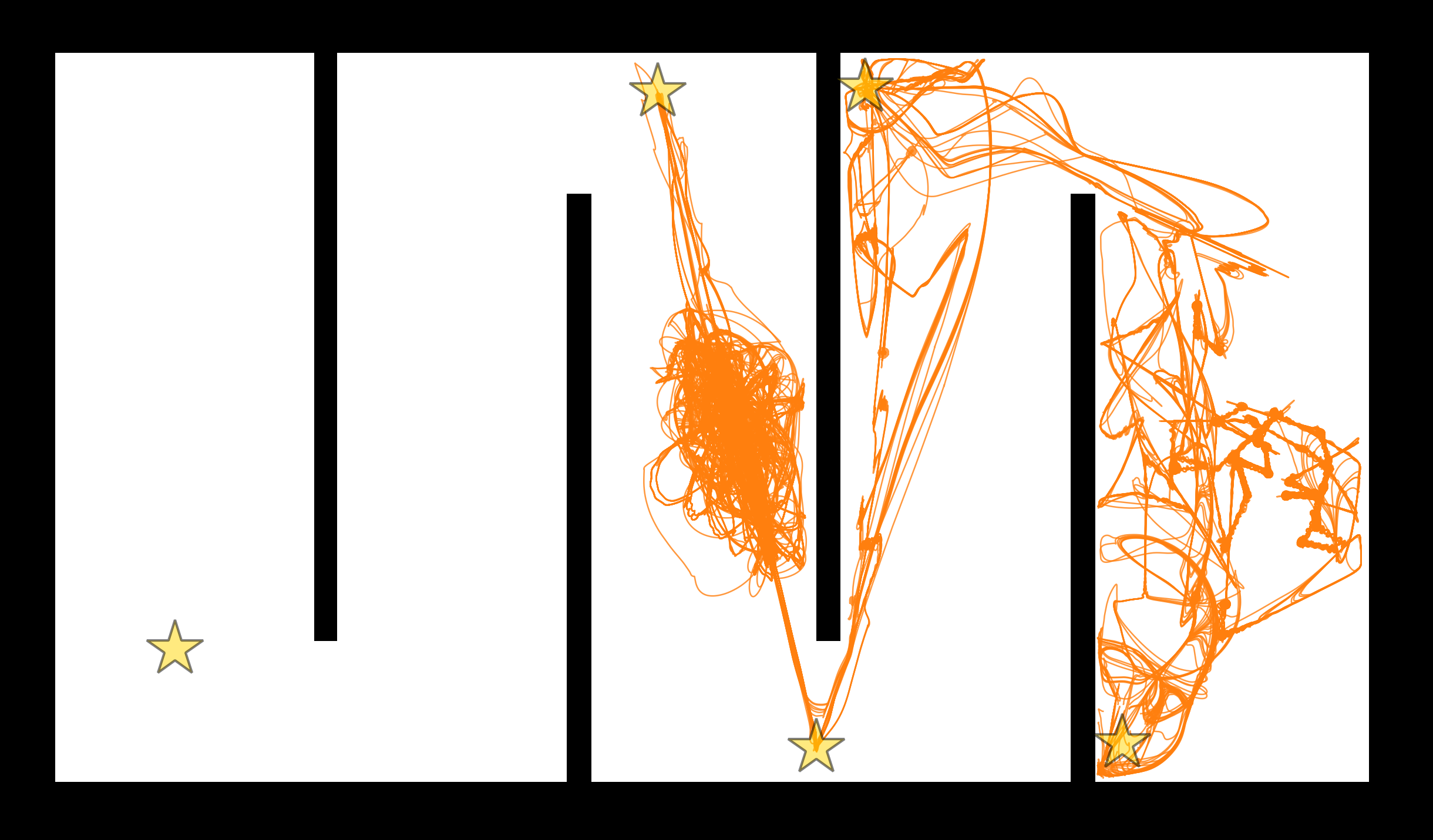}
        \label{fig:default_hairpin_traceplot_r5_after}
    \end{subfigure}
    \begin{subfigure}{0.336\textwidth}
        \centering
        \vspace{-0.2cm}
        \caption{R1 Capture $t \leq 8.95$s}
        \vspace{-0.2cm}
        \includegraphics[width=\linewidth]{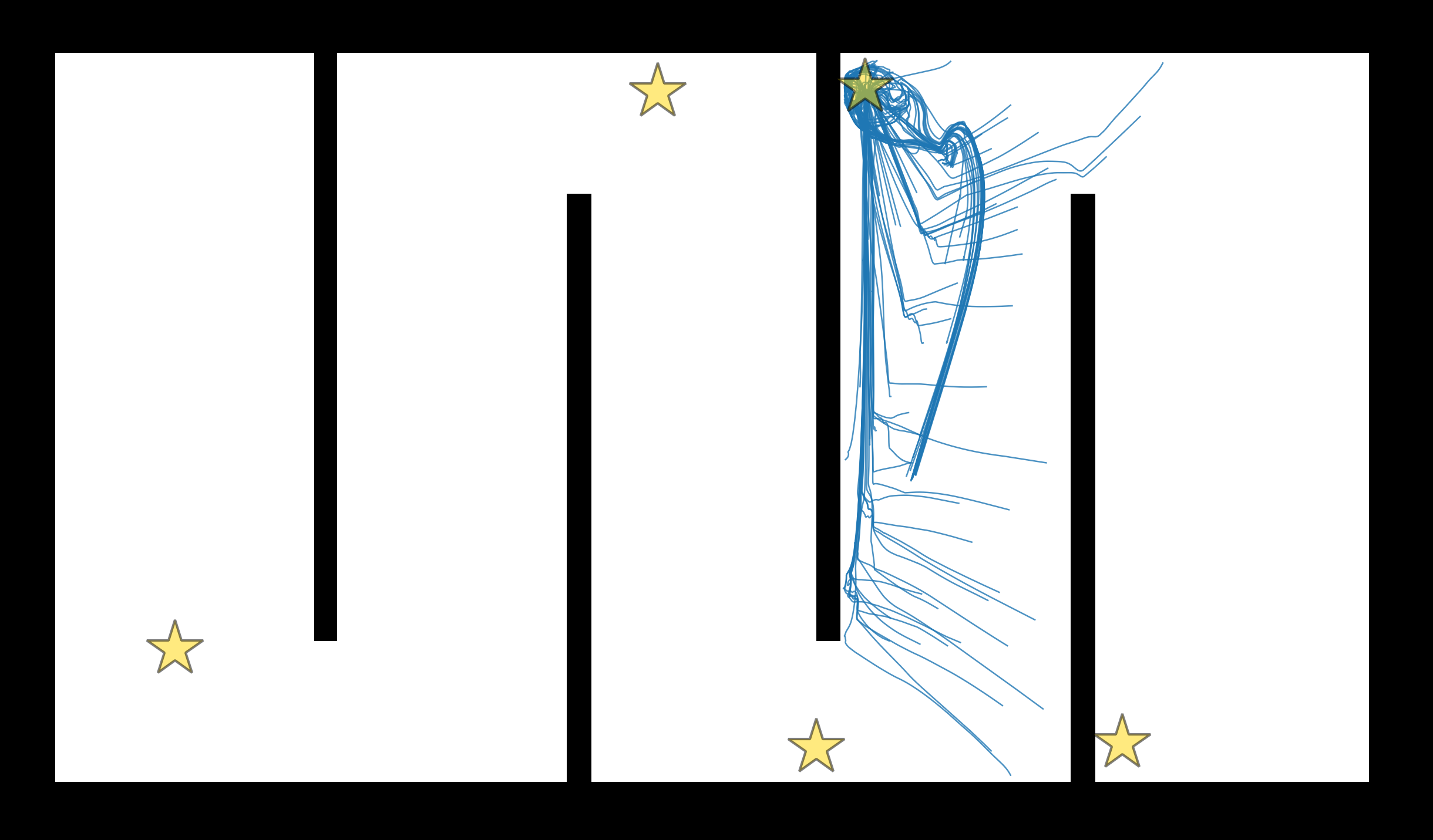}
        \label{fig:default_hairpin_traceplot_r1_capture}
    \end{subfigure}
    \begin{subfigure}{0.336\textwidth}
        \centering
        \vspace{-0.2cm}
        \caption{After R1 Capture $t > 8.95$s}
        \vspace{-0.2cm}
        \includegraphics[width=\linewidth]{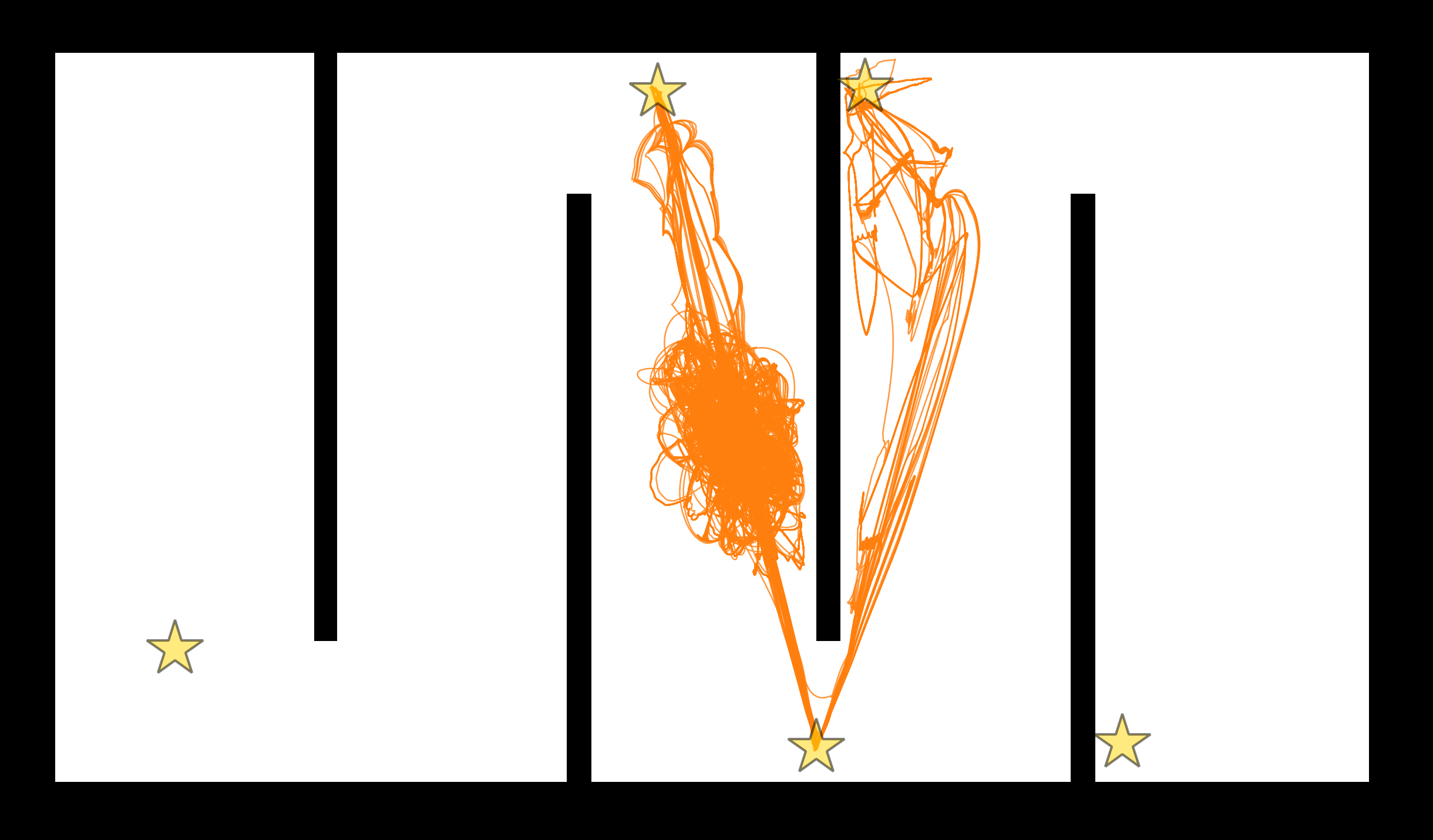}
        \label{fig:default_hairpin_traceplot_r1_after}
    \end{subfigure}
    \begin{subfigure}{0.336\textwidth}
        \centering
        \vspace{-0.2cm}
        \caption{R4 Capture $t \leq 37.86$s}
        \vspace{-0.2cm}
        \includegraphics[width=\linewidth]{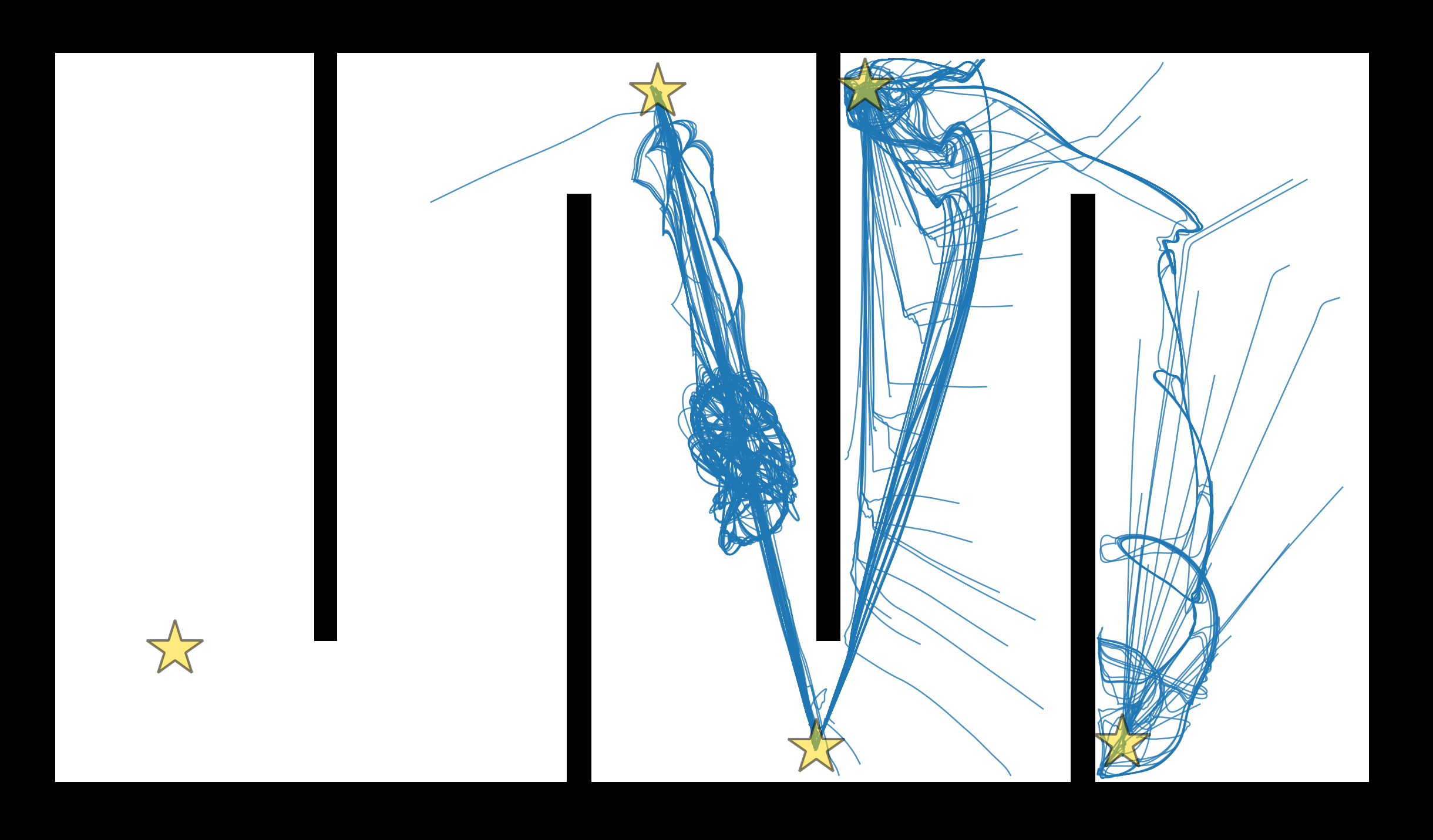}
        \label{fig:default_hairpin_traceplot_r4_capture}
    \end{subfigure}
    \begin{subfigure}{0.336\textwidth}
        \centering
        \vspace{-0.2cm}
        \caption{After R4 Capture $t > 37.86$s}
        \vspace{-0.2cm}
        \includegraphics[width=\linewidth]{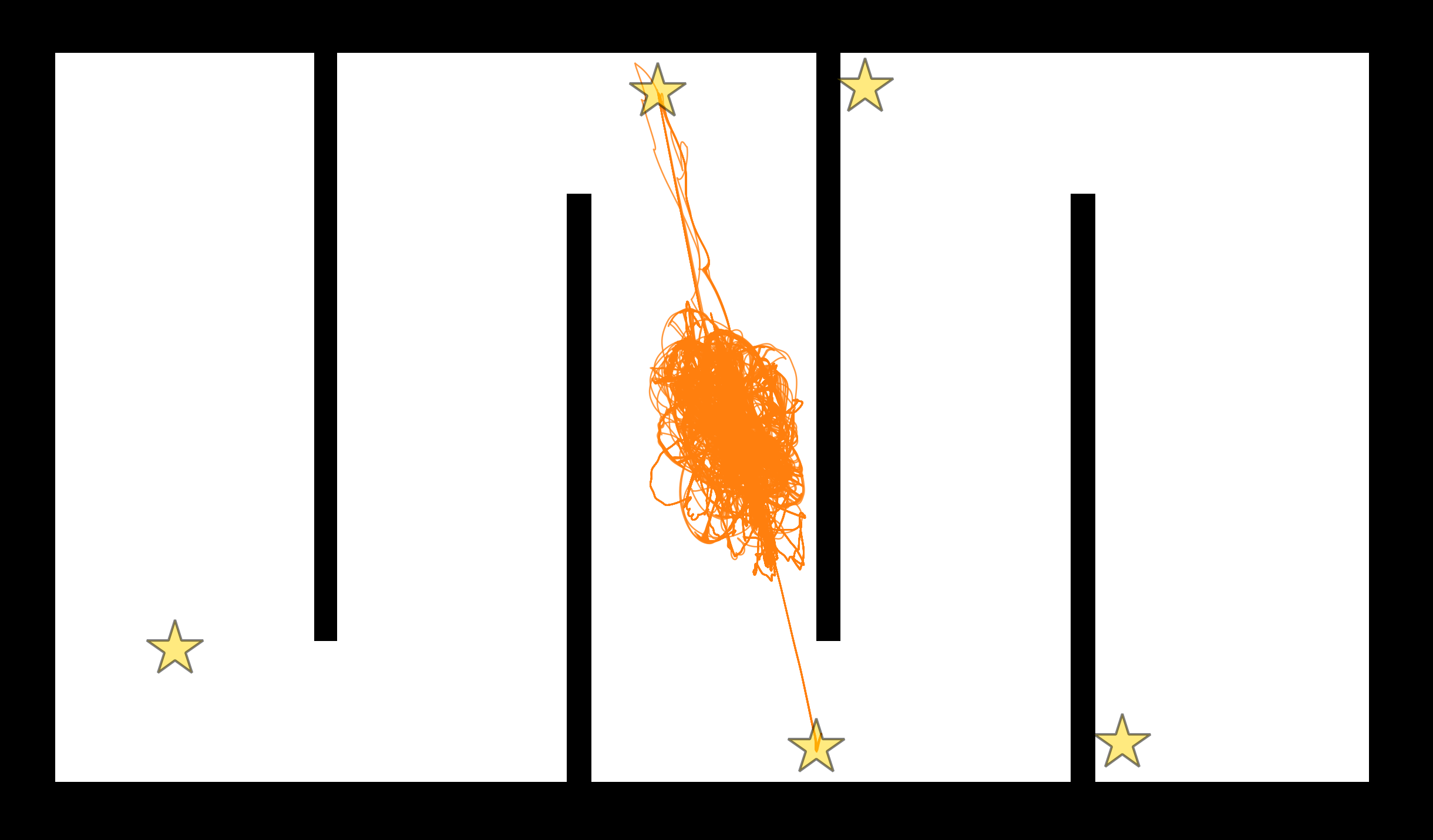}
        \label{fig:default_hairpin_traceplot_r4_after}
    \end{subfigure}
    \begin{subfigure}{0.336\textwidth}
        \centering
        \vspace{-0.2cm}
        \caption{R2 Capture $t \leq 41.02$s}
        \vspace{-0.2cm}
        \includegraphics[width=\linewidth]{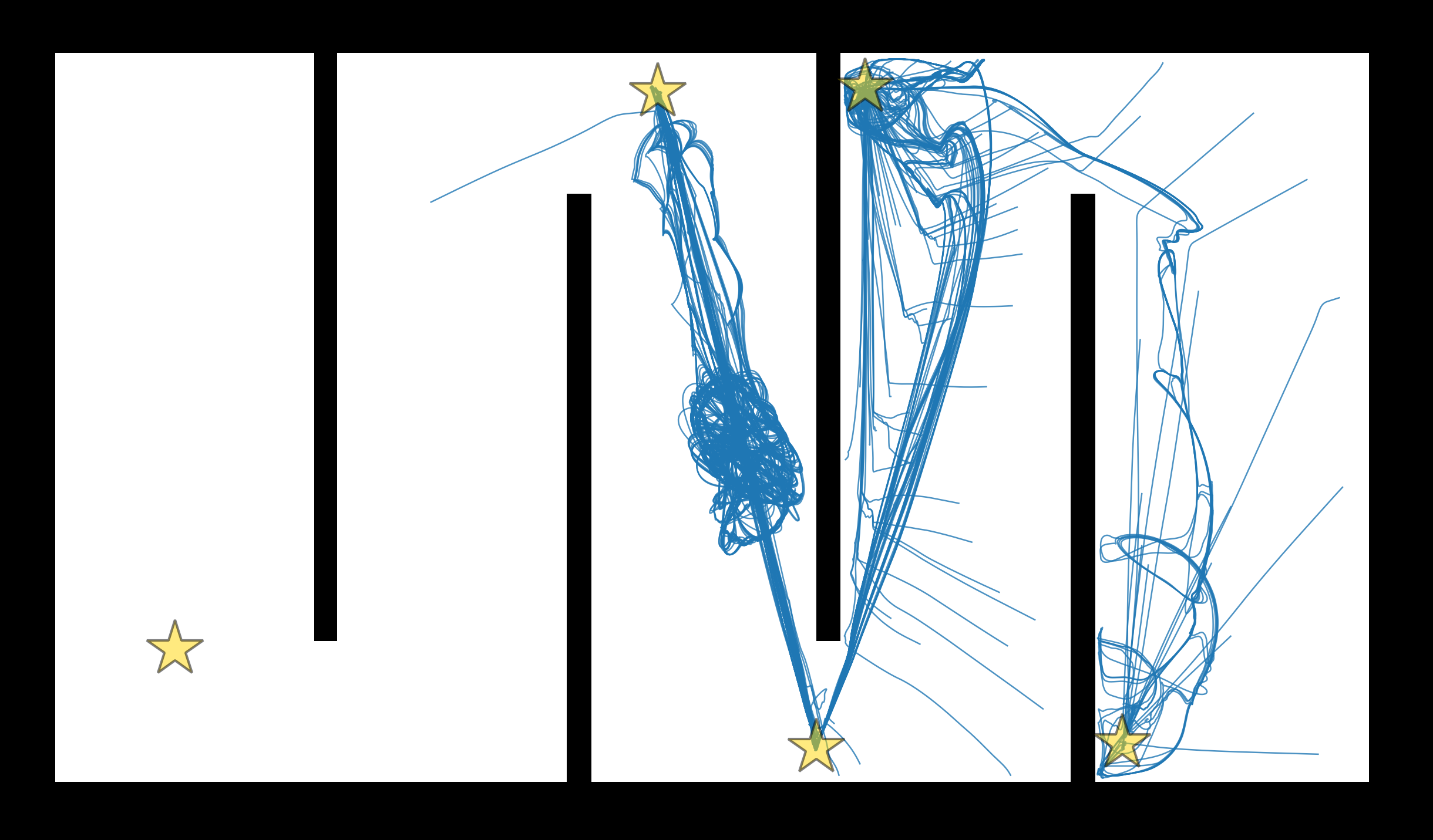}
        \label{fig:default_hairpin_traceplot_r2_capture}
    \end{subfigure}
    \begin{subfigure}{0.336\textwidth}
        \centering
        \vspace{-0.2cm}
        \caption{After R2 Capture $t > 41.02$s}
        \vspace{-0.2cm}
        \includegraphics[width=\linewidth]{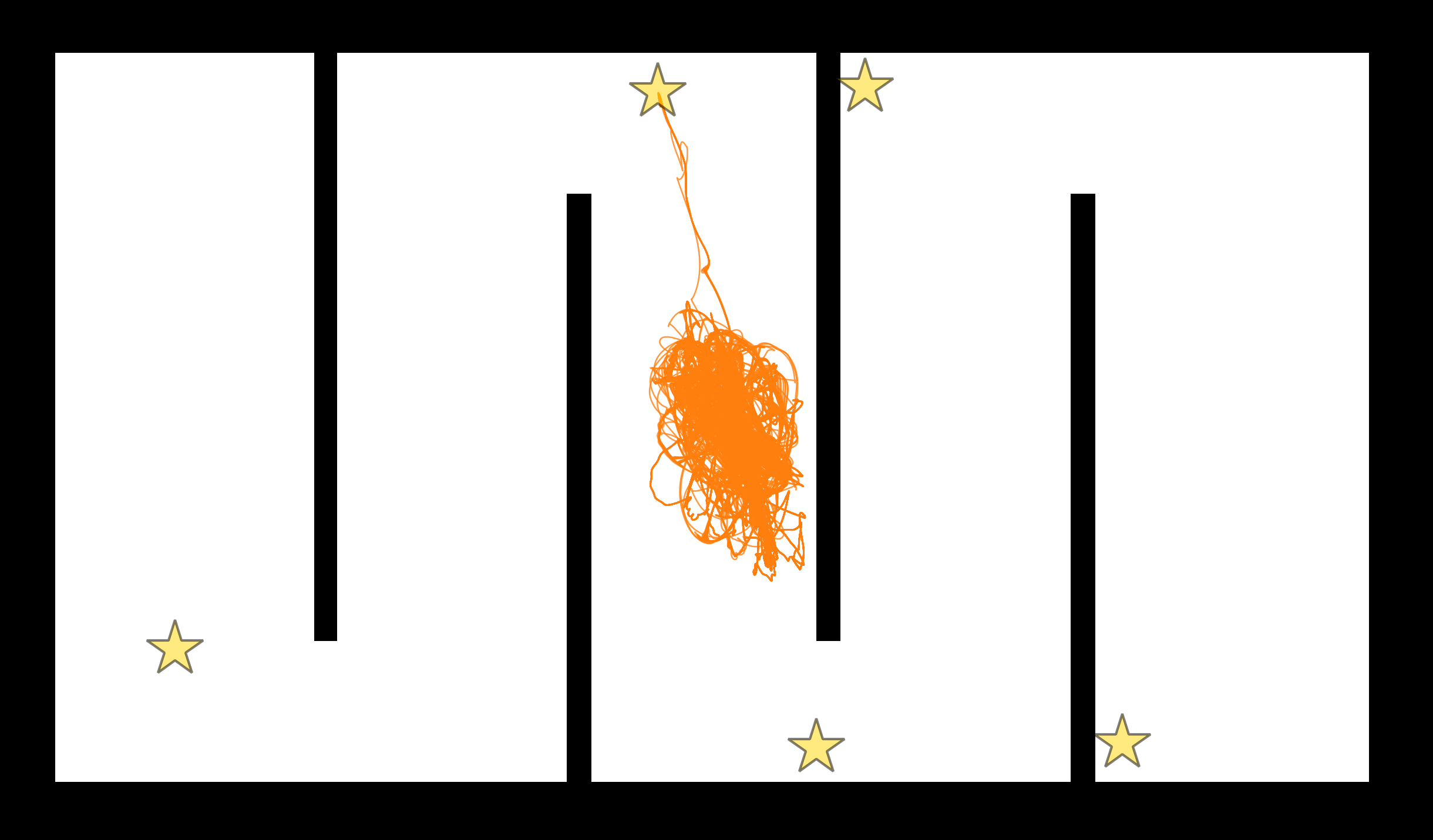}
        \label{fig:default_hairpin_traceplot_r2_after}
    \end{subfigure}
    \vspace{-0.5cm}
    \caption{Swarm trajectory trace in the Hairpin environment using default
parameters from~\cite{monaco2020cognitive} for said environment. Rewards,
denoted as R1-R5, are distributed in a counter clockwise fashion starting with
the top right-most star.}
    \label{fig:default_hairpin_traceplots}
\end{figure}

\begin{figure}[tb!]
    \centering
    \begin{subfigure}{0.36\textwidth}
        \centering
        \vspace{-0.2cm}
        \caption{R2 Capture $t \leq 2.02$s}
        \vspace{-0.2cm}
        \includegraphics[width=\linewidth]{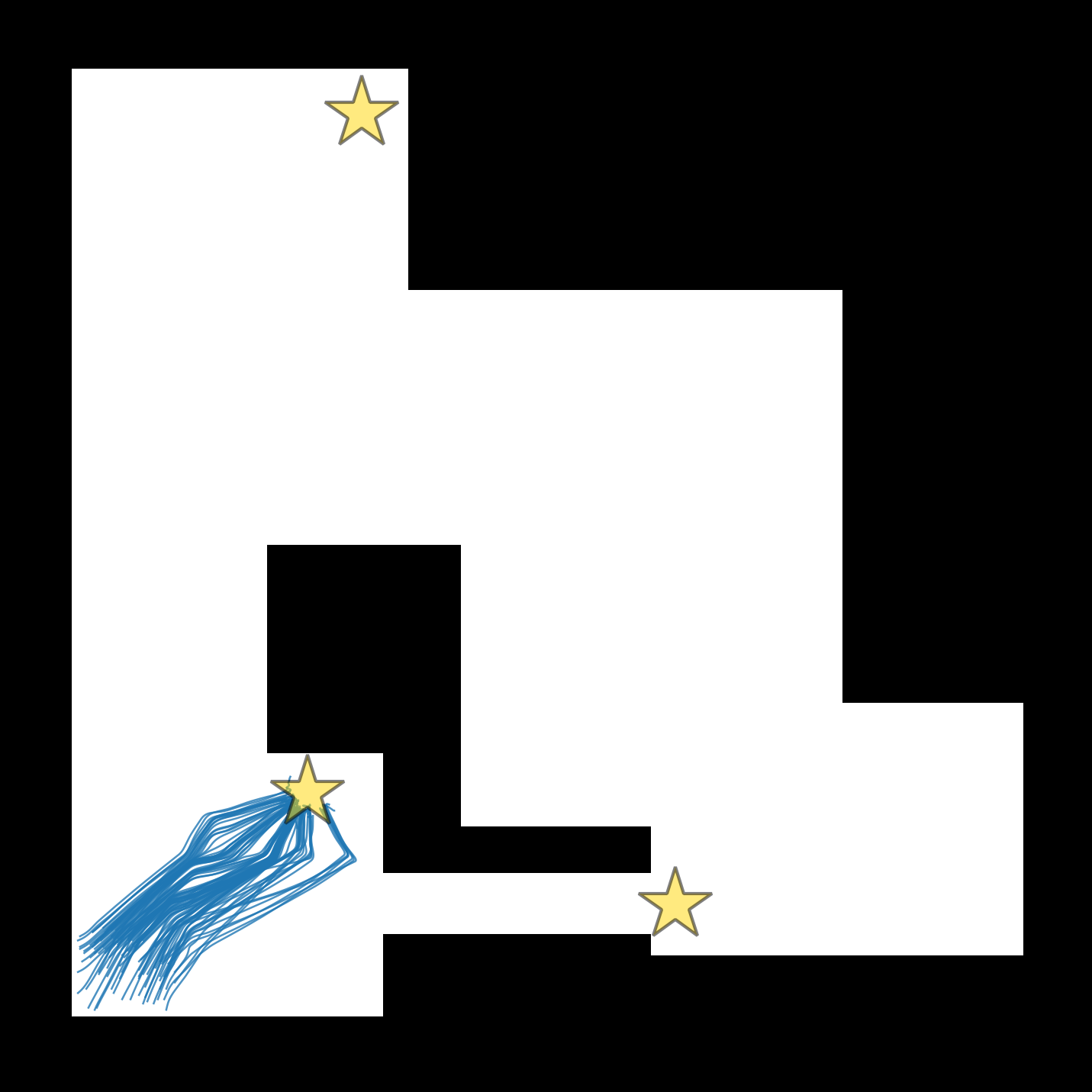}
        \label{fig:default_tunnel_traceplot_r2_capture}
    \end{subfigure}
    \begin{subfigure}{0.36\textwidth}
        \centering
        \vspace{-0.2cm}
        \caption{After R2 Capture $t > 2.02$s}
        \vspace{-0.2cm}
        \includegraphics[width=\linewidth]{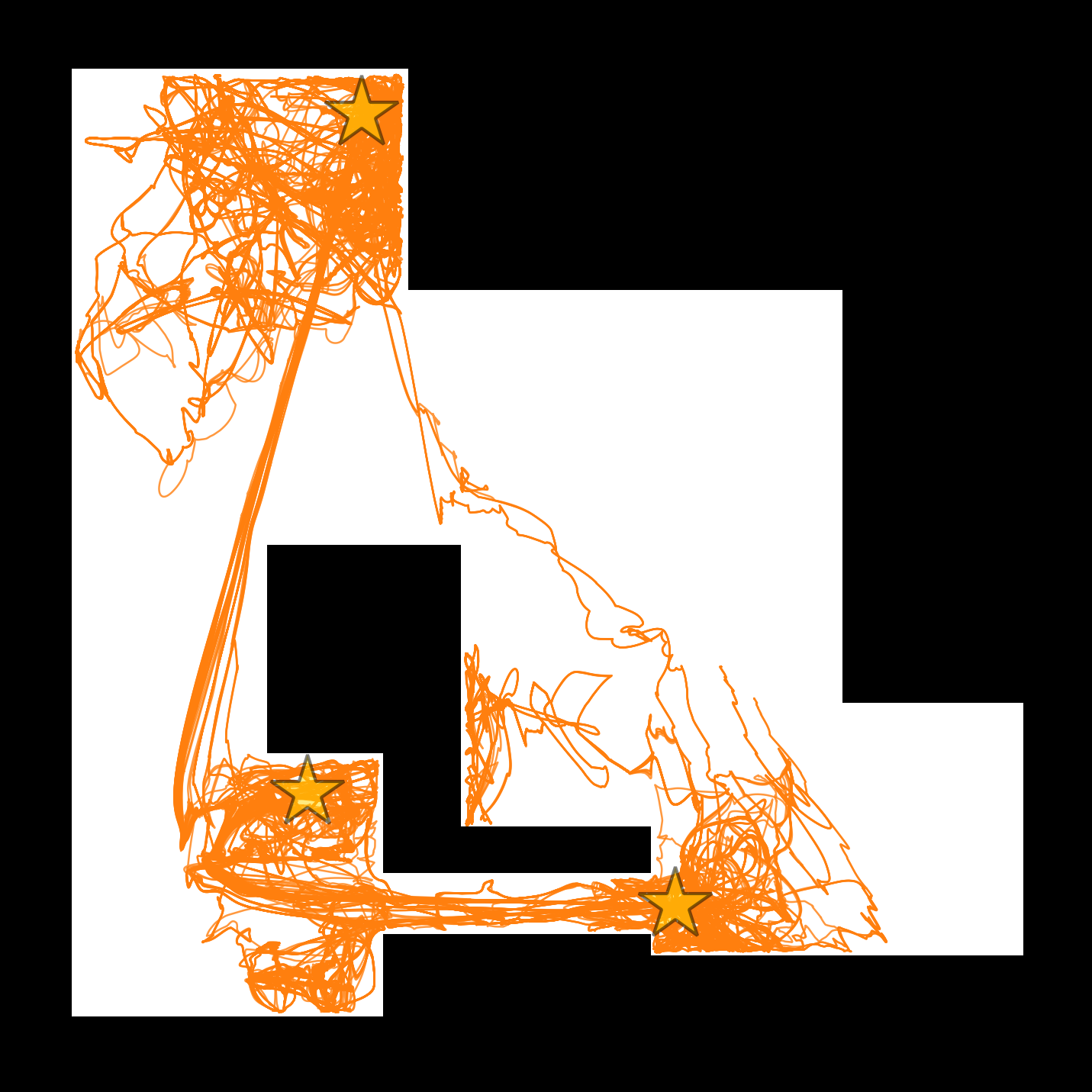}
        \label{fig:default_tunnel_traceplot_r2_after}
    \end{subfigure}
    \begin{subfigure}{0.36\textwidth}
        \centering
        \vspace{-0.2cm}
        \caption{R3 Capture $t \leq 34.88$s}
        \vspace{-0.2cm}
        \includegraphics[width=\linewidth]{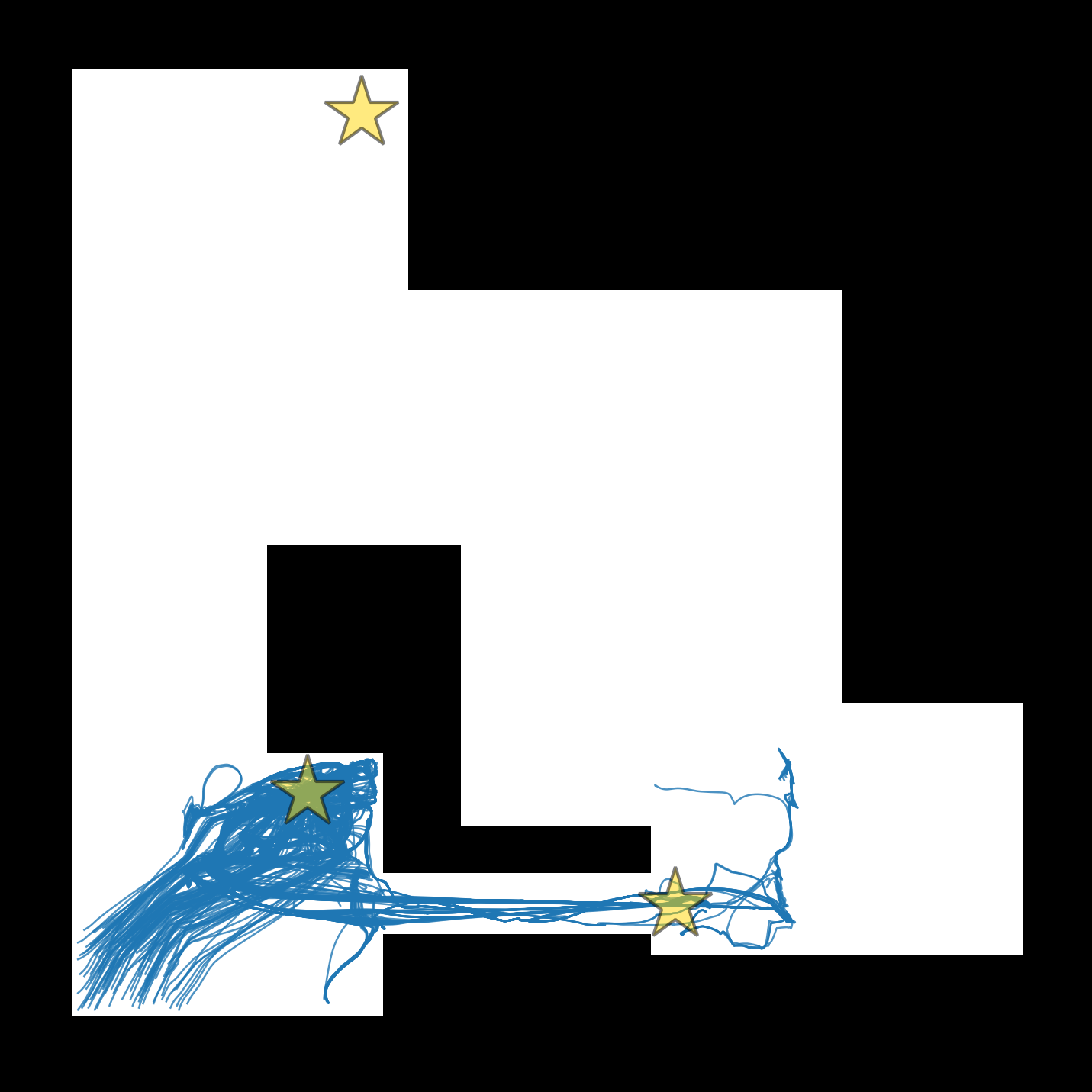}
        \label{fig:default_tunnel_traceplot_r3_capture}
    \end{subfigure}
    \begin{subfigure}{0.36\textwidth}
        \centering
        \vspace{-0.2cm}
        \caption{After R3 Capture $t > 34.88$s}
        \vspace{-0.2cm}
        \includegraphics[width=\linewidth]{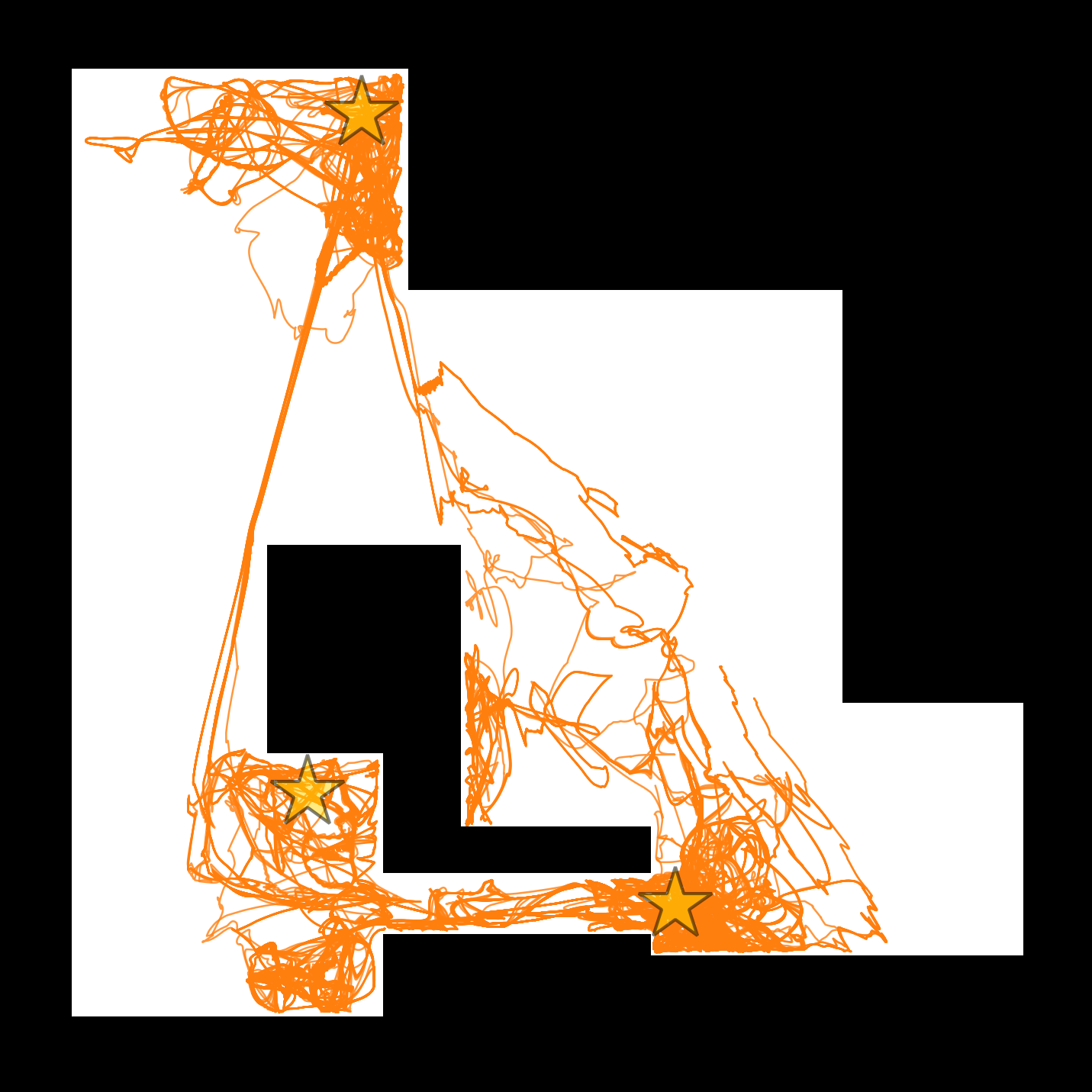}
        \label{fig:default_tunnel_traceplot_r3_after}
    \end{subfigure}
    \begin{subfigure}{0.36\textwidth}
        \centering
        \vspace{-0.2cm}
        \caption{R3 Capture $t \leq 175.42$s}
        \vspace{-0.2cm}
        \includegraphics[width=\linewidth]{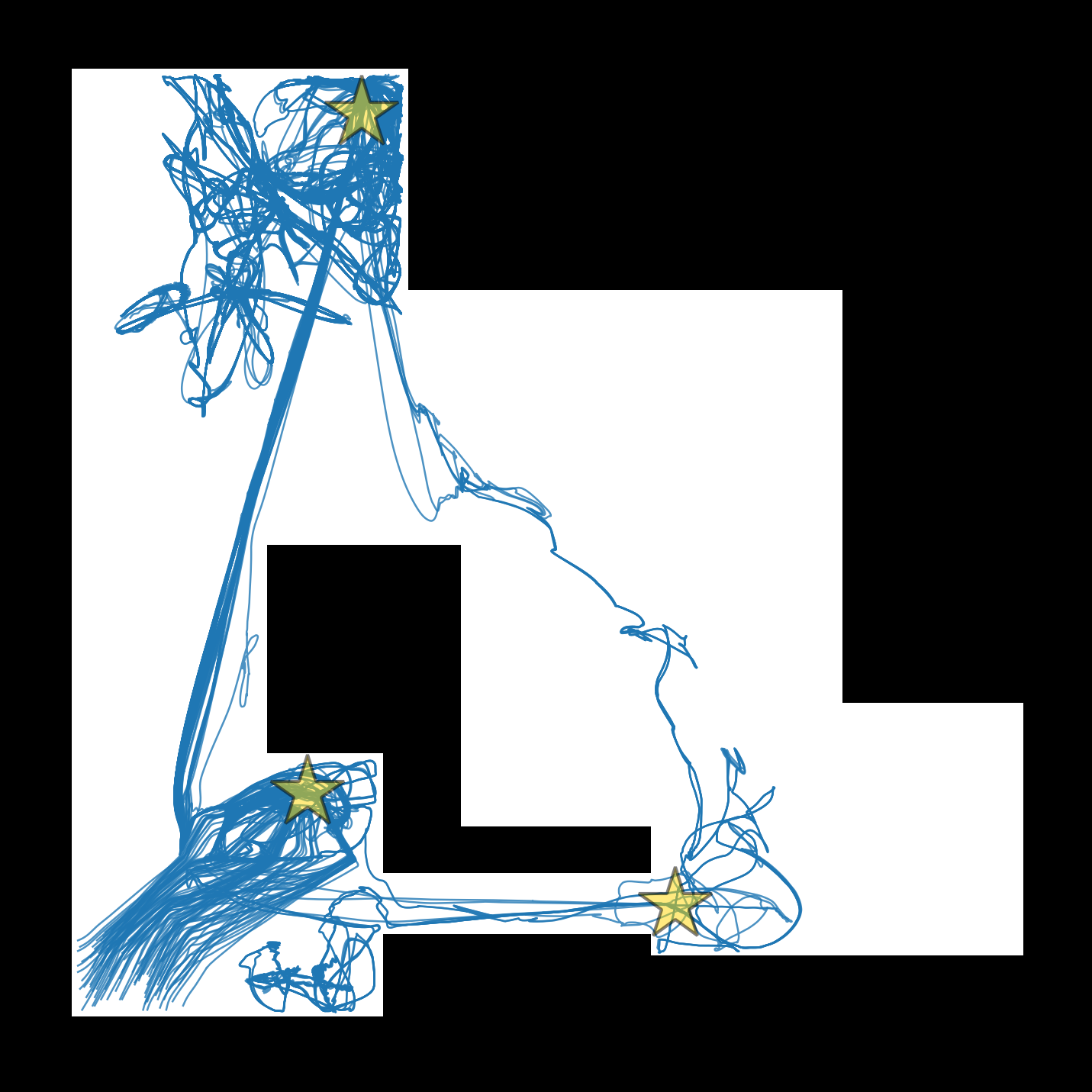}
        \label{fig:default_tunnel_traceplot_r1_capture}
    \end{subfigure}
    \begin{subfigure}{0.36\textwidth}
        \centering
        \vspace{-0.2cm}
        \caption{After R3 Capture $t > 175.42$s}
        \vspace{-0.2cm}
        \includegraphics[width=\linewidth]{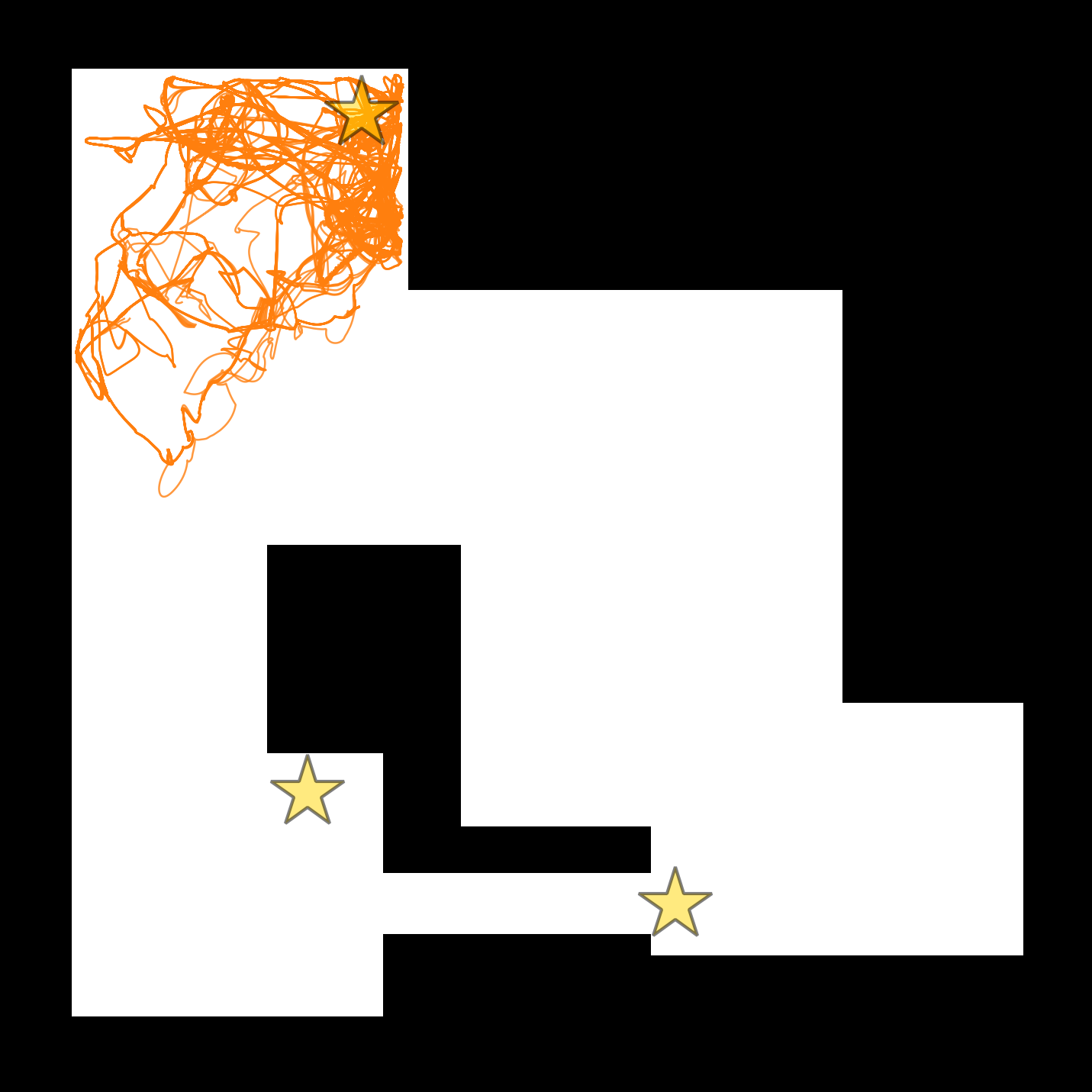}
        \label{fig:default_tunnel_traceplot_r1_after}
    \end{subfigure}
    \vspace{-0.5cm}
    \caption{Swarm trajectory trace in the Hairpin environment using default
parameters from~\cite{monaco2020cognitive} for said environment. Rewards,
denoted as R1-R3, are distributed in a counter clockwise fashion starting with
the top left-most star.}
    \label{fig:default_tunnel_traceplots}
\end{figure}

In Section~\ref{sec:experiment_results}, we identified that the default set
of parameters from Monaco et~al. (2020)~\cite{monaco2020cognitive} did not
perform as well as the acquisition functions-based \ac{gp} models. We showed, in
Figure~\ref{fig:obs_histogram}, the superior objective function loss performance
of the \ac{ei} and \ac{nei} the acquisition functions-based \ac{gp} models over
the default parameters. To further illustrate this discrepancy in fast reward
capture performance, we ran a simulation using the set of parameters from Monaco
et~al. (2020) specific to the hairpin environment, shown as trajectory plots in
Figure~\ref{fig:default_hairpin_traceplots}, and then repeated this process for
the tunnel environment (see Figure~\ref{fig:default_tunnel_traceplots}).

The manually-tuned default parameters poor objective function loss performance
can be attributed to the default swarm's slow reward-capture. The \ac{ei}-tuned
swarms was able to capture all five rewards on the hairpin environment
in 25.38s, as shown in Figure~\ref{fig:hairpin_traceplot_r2_capture},
where as the sluggish manually-tuned default swarm, shown in
Figure~\ref{fig:default_hairpin_traceplot_r2_capture}, took 41.02s. The
default parameters favor very high levels of reward exploration with a
$\kappa=6.6$, but low spatial exploration ($\sigma=2.0$) and equally low
inter-agent dynamics and reward response ($\eta_s = \eta_r = 1.0$). This
combination of parameters encourages exploration but poor cooperation,
which drastically increased the time-to-capture for all five rewards.
This slow reward capture behavior is further exacerbated on the tunnel
environment, shown in Figure~\ref{fig:default_tunnel_traceplot_r1_capture},
where the default parameter swarm took 175.42s to capture all three
rewards. For comparison, the \ac{ei}-tuned swarm was able to capture
all three rewards, see Figure~\ref{fig:tunnel_traceplot_r3_capture},
faster than the default swarm could capture two rewards (34.88s from
Figure~\ref{fig:default_tunnel_traceplot_r3_capture}). Furthermore, a 175.42s
all reward capture time is not even permitted for any of the \ac{gp}-based
methods as training criteria cuts off simulations at 120s. If maximum simulation
time limits were imposed on the default model, it would have only capture two
of the three rewards. Overall, these results illustrated the importance of
utilizing automated parameter tuning approaches like \ac{bo} for parameter
optimization instead of depending on default parameters used by prior works.

\end{appendices}

\appendix

\end{document}